\documentclass[]{interact}

\usepackage{epstopdf}
\usepackage{subfigure}
\usepackage{multirow}
\usepackage{geometry}
\usepackage[longnamesfirst,sort]{natbib}
\usepackage[colorlinks]{hyperref}
\usepackage{blindtext}

\usepackage{booktabs}
\usepackage{graphicx}
\usepackage[usenames, dvipsnames]{color}
\usepackage{natbib}
\bibpunct[, ]{(}{)}{;}{a}{,}{,}
\theoremstyle{plain}

\theoremstyle{definition}

\theoremstyle{remark}

\begin{document}
\title{Profile control chart based on maximum entropy}
\author{
\name{Seyedeh Azadeh Fallah Mortezanejad$^1$, Wang Ruochen$^2$ \thanks{CONTACT Wang Ruochen Email: wrc@ujs.edu.cn} Gholamreza Mohtashami Borzadaran$^3$, Renkai Ding$^4$, Kim Phuc Tran$^{5,6}$
}
\affil{$^{1,2,4}$ School of Automotive and Traffic Engineering, Jiangsu University, Zhenjiang, Jiangsu, China.\\
$^{3}$ Department of Statistics, Faculty of Mathematical Sciences Ferdowsi University of Mashhad, Mashhad, Iran.\\
$^{5}$Univ. Lille, ENSAIT, ULR 2461 - GEMTEX - Génie et Matériaux Textiles, F-59000 Lille, France\\
$^{6}$International Chair in DS \& XAI, International Research Institute for Artificial Intelligence and Data Science, Dong A University, Danang, Vietnam.}
}

\maketitle
\begin{abstract}
Monitoring a process over time is so important in manufacturing processes to reduce wastage in money and time. Some charts as Shewhart, CUSUM, and EWMA are common to monitor a process with a single intended attribute which is used in different kinds of processes with various ranges of shifts. In some cases, the process quality is characterized by different types of profiles. The purpose of this article is to monitor profile coefficients instead of a process mean. In this paper, two methods are proposed for monitoring the intercept and slope of the simple linear profile, simultaneously. In this regard, two methods are compared here. The first one is the linear regression, and another one is the maximum entropy principle. The $T^2$-Hotelling statistic is used to transfer two coefficients to a scalar. A simulation study is applied to compare the two methods in terms of the second type of error and average run length. Finally, two real examples are presented to demonstrate the applicability of the proposed chart. The first one is about semiconductors, and the second one is about pharmaceutical production processes. The performance of methods is relatively similar. The maximum entropy plays an important role in correctly identifying differences in the pharmaceutical example, while linear regression did not correctly detect these changes.
\end{abstract}


\section{Introduction}
Control charts have been widely used to monitor industry and manufacturing processes. Much researches have been done on how to monitor and detect shifts in the process or system, which is important to have suitable supervision on equipment performance. Kopnov and Kanajev $(1994)$ studied on burn breakdown of a special process and obtained an optimum limit respect to the existing costs on a degradation process. Haworth $(1996)$ showed how to use multivariate regression control charts to manage the maintenance process of the intended software. Xie et al. $(2000)$ published a paper on controlling the process reliability and mentioned on weaknesses and strengths of some control charts such as Shewhart. Chakraborti et al. $(2008)$ used joint statistical distributions in phase $ I $ to control the probability of earthquakes hazard and gained the probability of at least one wrong alarm during designing the control charts. They used two methods. In the first one, a certain amount was considered for the probability of false alarm, based on the Hiller $(1969)$ and Yang and Hiller $(1970)$ scheme. The second one concerned by King $(1954)$ scheme based on multivariate dependence distribution to get true and false alarms probabilities. Champ and Woodall $(1987)$ presented a simple and efficient method, based on using the Markov chain, and obtained the exact run length properties in Shewhart control charts.\\
CUSUM control charts have been generally used to detect smooth mean deviations. Westgard et al. $(1977)$ have worked on a control chart combination of Shewhart and CUSUM such that both rules are simultaneously employed and illustrated its better performance by a computer simulation study. It should be noted that CUSUM control charts are suitable for predictable shifts. Sparks (2000) designed an adaptive CUSUM (ACUSUM) control chart that is efficient at signaling a range of future expected but unknown changes.\\
Quality control charts based on EWMA are popular tools in monitoring processes parameter shifts. Neubauer $(1997)$ explained the basics of these chart types. Zou and Tsung $(2011)$ and Prabhu and Runger $(1997)$ presented some examples of the multivariate EWMA control chart. When some different attributes of a process are simultaneously intended to monitor, one of the most common control charts is $ T^2 $-Hotelling. Aparisi $(1996)$ expressed the $ T^2 $-Hotelling plot as an easily used chart and calculated a comparative control table of $ T^2 $. Aparisi and Haro $(2001)$ provided a way that increases the power of the test inthe $ T^2 $-Hotelling charts so that the shifts are detected faster than the other way. Alfaro and Ortega $(2008)$ introduced a more powerful plot than $ T^2 $-Hotelling. Generally, $ T^2 $-Hotelling control charts are an extension of univariate Shewhart control charts. \\
Profiles are different models applied in the cases that the products or process quality is characterized by a functional relationship between a response (dependent) and one or more explanatory (independent) variables. Woodall et al. $(2004)$ reviewed on how to monitor profiles in manufacturing processes. Woodall $(2007)$ has presented a general framework to control processes based on the profiles. Smith and Livesey $(1992)$ and Quercia et al. $(2012)$ studied the maximum entropy and profile. Krogh and Mitchison $(1995)$ worked on a profile family of DNA that had different disordered subfamilies and solved the optimum problem by maximum entropy principle. \\
In this paper, we are working on a new method of monitoring a production process instead of monitoring its corresponding mean. Therefore, we apply a simple linear profile for this aim. The simple linear profile is a functional relationship between a response and one explanatory variable, such a linear regression model. This paper aims to find out suitable control limits for the simple linear profile coefficients. One way of estimating is by applying the regular regression parameter estimations. Besides, we engaged one more method of estimating parameters named as maximum entropy principle. The basic step in maximum entropy is to get the unknown distribution function of an available dataset. The final result is called maximum entropy distribution. In the end, the unknown coefficients are approximated via this specified distribution. In most cases in the real world, the distribution function of a dataset is passive and has to be estimated beneficially. So, we explain it while describing our method in approximating the profile coefficients. In the end, some simulation examples and real data studies are added to make a good comparison between the traditional and the maximum entropy methods. \\
The structure of the rest of the present paper is as follows: In section \ref{ME}, Shannon entropy and maximum entropy distribution are rendered, because they are needed to compute the estimated values of profile coefficients. Some literature and basic definitions of  profiles are provided in Section \ref{profile}. In section \ref{CalcuCO}, the maximum entropy methods and the linear regression $ (LR) $ are applied to estimate the coefficients of a simple linear profile, and control limits based on the $ T^2 $-Hotelling statistic for two methods are obtained, which are adequate in detecting inappreciable shifts. Furthermore, a simulation study is done in section \ref{5}. Section \ref{6} presents an example of real data on the semiconductor production process and compares two proposed methods in the case. Section \ref{7} consists of a real pharmaceutical example, and the maximum entropy principle's power is demonstrated. In the end, conclusions are available in the last section.

\section{\bf Bivariate maximum entropy distribution} \label{ME}
Shannon entropy, introduced by Shannon $(1948)$, has been widely used as an information criterion to obtain the probability density function based on known knowledge of the available dataset applied in some fields of studies like computer science, physics, and economics, etc. For example, the distributions of financial variables, stock returns, and incomes can be gained by the maximum entropy concept just by defining some intentional constraints. The manner of estimating the maximum entropy distribution was presented first by Jaynes $(1957)$. For more information, see Zellner and Highfield $(1988)$, Wu $(2003)$, Wu and Perloff $(2007)$, and Chu and Satchell $(2009)$. Others like Kagan et al. $(1973)$ and Shore and Johnson $(1980)$. Some more papers on the multivariate states were reproduced such as Jones $(1976)$, Urzúa $(1988)$, Costa et al. $(2003)$, Kouskoulas et al. $(2004)$, Bhattacharya $(2006)$, Ebrahimi et al. $(2008)$, Pougaza and Djafari $(2011)$, Pougaza and Djafari $(2012)$, and Mortezanejad et al. $ (2019) $. For example, Fallah et al. $ (2019) $ worked on a capability index in the manufacturing process whose distribution has been obtained via maximum entropy principle.\\
First of all, to determine bivariate maximum entropy, let $ X $ and $ Y $ be two random variables whose joint density and distribution functions are $ f_{X,Y}(x,y) $ and $ F_{X,Y}(x,y)$, respectively.  Their joint Shannon entropy is as below:
\begin{equation*}
  H(f)=-\int\int_{\mathbb{S}(X,Y)}\log f_{X,Y}(x,y)~ dF_{X,Y}(x,y),
\end{equation*}
where $ \mathbb{S}(X,Y) $ is the joint support set of $ X $ and $ Y $, and $$ dF_{X,Y}(x,y)=\frac{\partial^2 F_{X,Y}(x,y)}{\partial x~ \partial y}~dx~dy=f_{X,Y}(x,y)~dx~dy. $$
Let's find the joint maximum entropy distribution of $ X $ and $ Y $ under these constraints based on mathematical expectation definitions:
\begin{eqnarray}\label{cons}
E(h_i(X,Y)|F) &=&\int\int_{\mathbb{S}(X,Y)}h_i(X,Y)~dF_{X,Y}(x,y) \\ \nonumber
&=& m_i(x,y),~~~i=1,~\ldots,~r,
\end{eqnarray}
where $r$ is the number of constraints, $ m_i(\cdot) $s are arbitrary moments functions toward $ i=1,~\ldots,~r $, for example, one choice of $m_i(x,y)$ can be $ \overline{xy}=n^{-1}\sum_{j=1} ^n x_i~y_i$ whose value is computed by the existence dataset, and $n$ is the size of each sample. In equation \eqref{cons}, $ x $ and $ y $ are the observations of $ X $ and $ Y $ variables, and $ F $ is their true joint distribution function. But, it does not determine, so our aim in this section is to approximate it via the maximum entropy principle. $ h_i(X,Y) $s for $ i=1,~\ldots,~r $ are intended functions according to $ m_i(\cdot) $s, for instance if $m_i(x,y)=\overline{xy}$, then its corresponding $h_i(X,Y)$ is $XY$. To approximate the unknown joint distribution function $ F $ or density function $ f $ via maximum entropy principle, we need to apply the Lagrange function based on Shannon entropy and some optional constraints in \eqref{cons} as follows:
\begin{eqnarray}\label{L}
L(f,\lambda_0,~\ldots,~\lambda_r)&=&-\int\int_{\mathbb{S}(X,Y)} \log f_{X,Y}(x,y) dF_{X,Y}(x,y)-\lambda_0 \{ \int\int_{\mathbb{S}(X,Y)} dF_{X,Y}(x,y)-1\}   \nonumber\\
&-&\Sigma_{i=1} ^r \lambda_i \{ \int\int_{\mathbb{S}(X,Y)} h_i(x,y) dF_{X,Y}(x,y)-m_i(x,y)\}. \nonumber
 \end{eqnarray}
The coefficient equation of $ \lambda_0 $ ensures that the result is a density function according to this fact:
\begin{equation}\label{denC}
  \int\int_{\mathbb{S}(X,Y)} dF_{X,Y}(x,y) =1.
\end{equation}
So by differentiation the Lagrange function respect to $ f $, we get an equation as a function of $f$:
\begin{equation*}
  \frac{\partial L(f,\lambda_0,~\ldots,~\lambda_r)}{\partial f}=-\log f -1-\lambda_0-\Sigma_{i=1} ^r \lambda_i ~ h_i(X,Y).
\end{equation*}
Then, the following equation has to be solved respect to $ f $:
\begin{equation*}
 \frac{\partial L(f,\lambda_0,~\ldots,~\lambda_r)}{\partial f}=0.
\end{equation*}
Finally, the density of maximum entropy is derived:
\begin{equation}\label{MEf}
 f_{X,Y}(x,y) = \exp (-1-\lambda_0-\Sigma_{i=1} ^r \lambda_i ~ h_i(x,y)),~~ (x,y)\in \mathbb{S}(X,Y).
\end{equation}
The Lagrange coefficients $\lambda_0,\ldots,\lambda_r$ have to be computed by substituting function \eqref{MEf} in desired constraints \eqref{cons} and \eqref{denC}, then the maximum entropy coefficients can be calculated by solving a system of equations respect to $ \lambda $s, and we will explain more in the following. In the next section, we intend to calculate the coefficients of the profile based on the maximum entropy principle and make a comparison with the linear regression method and its coefficients.

\section{\bf Profiles definition} \label{profile}
The profile consists of a response variable and one or more independent or explanatory variables that are used to review and monitor a manufacturing process over time. It can be presented as a simple linear regression such as Mahmoud and Woodall $(2004)$, and Gupta et al. $(2006)$ reported. Kang and Albin $(2000)$ have expressed two techniques to monitor a multivariate profile by the $ T^2 $-Hotelling control chart. Mahmoud et al. $(2007)$ have presented a manner based on likelihood ratio statistics to analyze the data of phase $ I $. In the literature of profiles, such papers as Quandt $(1958)$, Holbert $(1982)$, Hawkins $(1989)$, Kim and Siegmund $(1989)$, Kim $(1994)$, and Chen $(1998)$ are available. They have worked on how to detect shifts or change points of processes using simple linear profiles. Lawless et al. $(1999)$ have given some examples of linear profiles in automotive engineering. Kang and Albin $(2000)$ have presented two examples of different situations of producing products with profiles. Some authors such as Mandel $(1969)$, Hawkins $(1991)$, Hawkins $(1993)$, Zhang $(1992)$, and Wade and Woodall $(1993)$ have monitored linear profiles via regression adjusted control charts. \\
Profiles have several types like simple linear profile, multivariate linear profile, multiple linear profile, polynomial linear profile, generalized linear profile, and nonlinear profile. However, in this paper, our focus is on a simple linear profile. First of all, suppose that $ n $ fixed points of observations are exist from variable $X$, $ k $ random samples are obtained from the response variable $Y$ over time. Also, assume that in controlled situation response or dependence variable $ Y $ and independent or explanatory variable $ X $ are modeled as below that is called simple linear profile:
\begin{equation}\label{ProfileR}
  \underset{\sim}{y_j}=a+b ~ \underset{\sim}{x}+\underset{\sim}{\varepsilon_{j}},~~j=1,~\ldots,~k,
\end{equation}
where $ k $ is the number of total samples  that each of them has size $ n $, $ \underset{\sim}{x} $ is the observations of the independent variable $X$, and intrinsic error $ \underset{\sim}{\varepsilon_{j}} $ is an independent random variable with a normal distribution of zero mean and fixed variance of $ \sigma^2 $. The intercept $a$ and slope $b$ in the model are named as profile coefficients.  The model \eqref{ProfileR} is similar to simple linear regression, but their discrepancy is in the witnessing vector of $X$. In the regression model, there are different vectors of $ \underset{\sim}{y_j} $ and $ \underset{\sim}{x_j} $ for each sample, and the model is $ \underset{\sim}{y_j}=a+b ~ \underset{\sim}{x_j}+\underset{\sim}{\varepsilon_{j}} $ for $j=1,~\ldots,~k$. But in the profile model, there are only different vectors of response like $ \underset{\sim}{y_j} $, and the same vector of $ \underset{\sim}{x} $ is used for different observation vectors of $Y $ as they can be seen in \eqref{ProfileR} where we have a fixed vector of $ \underset{\sim}{x} $ for each $ k $ sample. \\
In this paper, our purpose is to monitor these profiles coefficients $a$ and $b$ of a process instead of monitoring the process mean the duration of passing the time, and we interest in observing possible shifts by the changes of the desirable coefficients.

\section{\bf Calculation of profile coefficients}\label{CalcuCO}
In this section, we present a new method of finding profiles coefficients based on the maximum entropy principle. The result is comparing with the linear regression coefficients. As we mentioned before, there are two coefficients $a$ and $b$. In the following,  we will define a two-dimension vector-like $ (a,~b)$. Therefore, we applied the $ T^2 $-Hotelling statistic to reduce the dimension to $ 1$. Then, we plot these values and calculate $ ARL_0 $ and $ ARL_1 $  in section \ref{5} for a range of small shifts in both manners. Here we have three shift types on intercept, slope, and simultaneous of both as a mixed model. Finally, their second type of error $ \beta $ is simultaneously plotted to judge them. Let $ \underset{\sim}{x}=(x_1,~\ldots,~x_n) $ be the fixed observations of $ X $. Here, we have $ k $ samples of dependence variable $ Y $ with length $ n $:
\begin{equation*}
  (\underset{\sim}{y_1},\underset{\sim}{x}),~\ldots,~(\underset{\sim}{y_k},\underset{\sim}{x}).
\end{equation*}
Now we would like to calculate the among of $ a $s and $ b $s in $ k $ samples via maximum entropy principle and linear regression. Hence, we desire to present some notations first such that:
\begin{displaymath}
\left\{\begin{array}{ll}
\underset{\sim}{m_1}=(\widehat{a}_{1-ME},\widehat{b}_{1-ME}),~\ldots,~\underset{\sim}{m_k}=(\widehat{a}_{k-ME},\widehat{b}_{1-ME}), \\
\underset{\sim}{l_1}=(\widehat{a}_{1-LR},\widehat{b}_{1-LR}),~\ldots,~\underset{\sim}{l_k}=(\widehat{a}_{k-LR},\widehat{b}_{k-LR}). \\
\end{array}\right.
\end{displaymath}
For instance, vector $ (\widehat{a}_{1-ME},\widehat{b}_{1-ME}) $ includes the estimated values of coefficients in \eqref{ProfileR} using the maximum entropy principle for the first sample, and we named this vector by $ \underset{\sim}{m_1} $. The same meaning is drawn from $ (\widehat{a}_{1-LR},\widehat{b}_{1-LR}) $ that is  called $ \underset{\sim}{l_1} $ with the difference of estimation procedure that is regression method or least square error. In the next step, we have to estimate the unknown distribution function of each sample to calculate the corresponding coefficients $ \underset{\sim}{m_1},~~\ldots,~\underset{\sim}{m_k} $. As mentioned in section \ref{ME}, to compute the maximum entropy distribution, some favorite constraints are needed. Here, we determine six constraints which are presented as follow:
\begin{equation}
 \begin{cases}
\int\int_{\mathbb{S}(X,Y)} dF_{X,Y}(x,y)=1, \\
\int\int_{\mathbb{S}(X,Y)} x dF_{X,Y}(x,y)=\overline{x}, \\
\int\int_{\mathbb{S}(X,Y)} y_j dF_{X,Y}(x,y)=\overline{y_j},~j=1,~\ldots,~k \\
\int\int_{\mathbb{S}(X,Y)} x^2 dF_{X,Y}(x,y)=\overline{x^2}, \\
\int\int_{\mathbb{S}(X,Y)} y_j ^2 dF_{X,Y}(x,y)=\overline{y_j ^2},~j=1,~\ldots,~k \\
\int\int_{\mathbb{S}(X,Y)} x y_j dF_{X,Y}(x,y)=\overline{x y_j},~j=1,~\ldots,~k. \\
  \end{cases} \label{constraints}
\end{equation}
Since our knowledge, the first equation is the guarantee for the result being a valid density function. So, we find the Lagrange coefficients $ \lambda_0,~\ldots,~\lambda_r $ in somehow that the final $ f_{X,Y}(\cdot,\cdot) $ be a density function. To do that, we have to institute the function of \eqref{MEf} in each equation of the system. Then, the system equations \eqref{constraints} has to be solved for Lagrange coefficients. It is obvious that the estimated joint distribution functions of $ X $ and $ Y $ are going to be obtained via maximum entropy concept for each sample $ (\underset{\sim}{y_j},\underset{\sim}{x}) $, $ j=1,~\ldots,~k $. Now we have $ k $ distribution functions according to each sample. $ a $ and $ b $ in maximum entropy principle are approximated as follows:
\begin{equation}\label{BhadM}
  \widehat{b}_{j-ME} = \frac{E_j[(X-EX)(Y-E_jY)]}{E[(X-EX)^2]},
\end{equation}
\begin{equation}\label{AhadM}
  \widehat{a}_{j-ME} = E_j(Y)-\widehat{b}_{j-ME} ~E(x),~ j=1,~\ldots,~k,
\end{equation}
where $ E_j(\cdot) $ and $ E(\cdot) $ are the mathematical expectation functions based on the maximum entropy estimation of density function $ f_{X,Y}(x,y) $. Corresponding linear regression coefficients are computed with $ k $ different samples too:
\begin{equation*}
  \widehat{b}_{j-LR}=\frac{\sum_{i=1} ^n (X_i-\overline{X})(Y_{ij}-\overline{Y}_j)}{\sum_{i=1} ^n (X_i-\overline{X})^2},
\end{equation*}
\begin{equation*}
  \widehat{a}_{j-LR}=\overline{Y}_j-\widehat{b}_{j-LR}~\overline{X},~ j=1,~\ldots,~k.
\end{equation*}
Our goal is to plot $ \underset{\sim}{m_1},~\ldots,~\underset{\sim}{m_k} $ and $ \underset{\sim}{l_1},~\ldots,~\underset{\sim}{l_k} $ to detect shifts of a process instead of monitoring its means. To do this, we use $ T^2 $-Hotelling statistic for all samples $ j=1,~\ldots,~k $ as below, and we name them $ T^2_{j-ME} $ and $ T^2_{j-LR} $ for the maximum entropy and regression methods, respectively:
\begin{equation}\label{T2fM}
  T^2_{j-ME}=(\underset{\sim}{m_j}-\underset{\sim}{\overline{m}})' S_m ^{-1} (\underset{\sim}{m_j}-\underset{\sim}{\overline{m}}),~ j=1,~\ldots,~k,
\end{equation}
\begin{equation}\label{T2fL}
    T^2_{j-LR}=(\underset{\sim}{l_j}-\underset{\sim}{\overline{l}})' S_l ^{-1} (\underset{\sim}{l_j}-\underset{\sim}{\overline{l}}),~ j=1,~\ldots,~k,
\end{equation}
where $ \underset{\sim}{\overline{m}} $ and $ \underset{\sim}{\overline{l}} $ are means related to $ \underset{\sim}{m_1},~\ldots,~\underset{\sim}{m_k} $ and $ \underset{\sim}{l_1},~\ldots,~\underset{\sim}{l_k} $, and also they are two-dimension like all $ \underset{\sim}{m_j} $ and $ \underset{\sim}{l_j} $. $ S_m $ and $ S_l $ are the estimations of the variance-covariance matrix of $ \underset{\sim}{m_1},~\ldots,~\underset{\sim}{m_k} $ and $ \underset{\sim}{l_1},~\ldots,~\underset{\sim}{l_k} $, respectively, because all $ \underset{\sim}{m_j} $ and $ \underset{\sim}{l_j} $ contain the estimation values of coefficients. Moreover $S_m ^{-1} $ and $ S_l ^{-1} $ are their corresponding matrix inverse. Then, we apply two different upper control limits $ (UCL) $ to check $ T^2 $s to be in statistical control. It is worth to mention that the lower control limits $ (LCL) $ are $ 0 $, because this statistic is always positive. The first $ UCL $ used here is based on Fisher distribution that is the same for all $ T^2 _{j-ME} $ and $ T^2_{j-LR} $, $(j=1,~\ldots,~k)$:
\begin{displaymath}
\left\{\begin{array}{ll}
UCL_F=\frac{p(k+1)(k-1)}{k^2-pk}F_{\alpha,p,k-p}, \\
LCL=0, \\
\end{array}\right.
\end{displaymath}
where $ p $ is the number of profile coefficients that are $ 2 $ here. If the number of samples $ k $ is more than $ 100 $, the control limits are changed to:
\begin{displaymath}
\left\{\begin{array}{ll}
UCL_F=\frac{p(k-1)}{k-p}F_{\alpha,p,k-p}, \\
LCL=0. \\
\end{array}\right.
\end{displaymath}
The second control limits are quantile of $ T^2 _{1-ME},~\ldots,~T^2 _{k-ME} $ and $ T^2 _{1-LR},~\ldots,~T^2 _{k-LR}$ . These control limits are different for the maximum entropy and linear regression methods:
\begin{displaymath}
\left\{\begin{array}{ll}
UCL_{ME}=q_{ME}, \\
UCL_{LR}=q_{LR}, \\
\end{array}\right.
\end{displaymath}
where $q_{ME}$ and $q_{LR}$ are $(1-\alpha)\%$ $T^2$-Hotelling quantile values of the maximum entropy and linear regression techniques, respectively, and $\alpha$ is the first type of error determined $0.05$ in this paper. The two methods are suitable for discovering soft changes as we pointed before. $ ARL_0 $ and $ ARL_1 $ are estimated for smooth shifts of intercepts and slopes in the simulation study section to illustrate this assertion.

\section{\bf Simulation study}\label{5}
\begin{figure}[!h]
\begin{center}
\centering \vspace*{-.5cm}
\includegraphics[width=90mm]{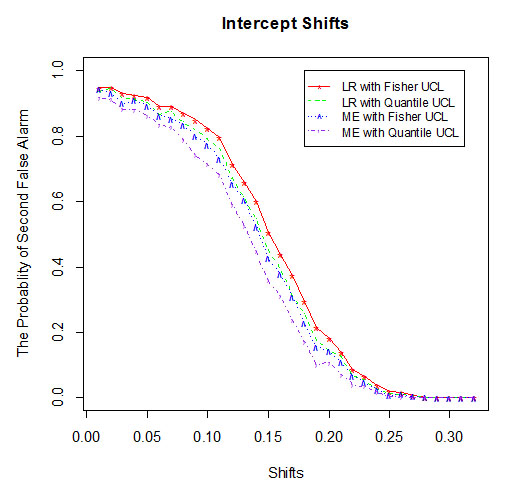}
\caption{The values of the second type of errors for the intercept shift in model $ I:~ Y=(2+s)+3X+\varepsilon $.}\label{1}
\end{center}
\end{figure}

\begin{figure}[!h]
\begin{center}
\centering \vspace*{-.5cm}
\includegraphics[width=90mm]{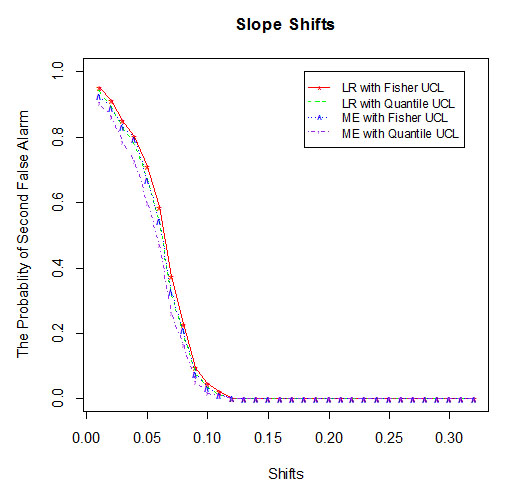}
\caption{The values of the second type of errors for the slope shift in model $ II:~ Y=2+(3+s)X+\varepsilon $.}
\label{2}
\end{center}
\end{figure}

\begin{figure}[!h]
\begin{center}
\centering \vspace*{-.5cm}
\includegraphics[width=90mm]{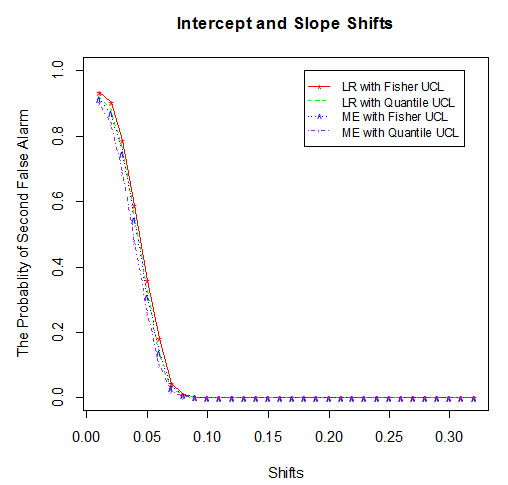}
\caption{The values of the second type of errors for the intercept and slope shifts in model $ III:~ Y=(2+s)+(3+s)X+\varepsilon $.}\
\label{3}
\end{center}
\end{figure}

In this section, we would like to use previous methods to find out shifts in a simulated dataset. Then, we would like to compare the abilities of the maximum entropy principle and the linear regression method. To do this, we first need to describe $ ARL_0 $ and $ARL_1 $ that are our tools to make a comparison. Average run length $ (ARL) $ of control charts is a suitable way to decide the sample size. $ ARL $ is the average number of in-control samples that have to be plotted before detecting an out of control sample. $ ARL_0 $ is usually calculated based on the first type of error $ \alpha $ for uncorrelated data as below:
\begin{equation*}
  ARL_0 = \frac{1}{\alpha},
\end{equation*}
where $ \alpha $ is the probability when an in-control sample reaches the control limits. $ ARL_0 $ is usually used to calculate the performance of control limits. $ ARL_1 $ is applied to discover shifts in the process means named as out of control $ ARL $ and defined as below:
\begin{equation*}
  ARL_1 =\frac{1}{1-\beta},
\end{equation*}
where $ \beta $ is the second type of error described as a probability of being in statistical control when a mean shift occurs. In the following section, we need to explain two different phases in controlling a process. Usually, the control checking of a process consisted of phase $ I $ and $ II $. The purpose and performances of these two phases are different, and researchers should pay attention to the differences. In phase $ I $, some pervious data is available under favorable conditions whose purpose is to get some information about the dispersion and stability of the process and to model its performance under desirable qualifications. Appropriate control limits are gained based on the available information. Then, it is checked whether the process has been in-control during this period or not and whether these available control limits are reliable for phase $ II $ or not. Furthermore, $ARL_0$ should be calculated in this phase too. In phase $ II $, there is some information from phase $I$ such that the process was in a stable state and the necessary parameters were estimated in phase $ I $. In the second phase, control limits of phase $ I $ are used to checking the process being in-control or not. The most important purpose of this phase is to detect faster any undesirable shifts or changes in the process parameters. Control charts in phase $ II $ are based on the $ ARL_1 $. Woodall $(2017)$ implied to disturbance of the theoretical and practical issues of statistical quality control in phases $ I $ and $ II $. Furthermore, how to collect information in phases $ I $ and $ II $ have been mentioned in the article. \\
Let's first look at a small simulation example and explain the steps to finding control limits using this article's methods. The simulated data example are in Table \ref{Sdata}. As mentioned before, some constraints are needed to find the maximum entropy distributions like \eqref{constraints}. For simplicity, the applied constraints for this example are:
\begin{equation}
 \begin{cases}
\int\int_{\mathbb{S}(X,Y)} dF_{X,Y}(x,y)=1, \\
\int\int_{\mathbb{S}(X,Y)} x dF_{X,Y}(x,y)=\overline{x}, \\
\int\int_{\mathbb{S}(X,Y)} y_j dF_{X,Y}(x,y)=\overline{y}_j,~j=1,~\ldots,~4 \\
\int\int_{\mathbb{S}(X,Y)} x y_j dF_{X,Y}(x,y)=\overline{x y}_j,~j=1,~\ldots,~4. \\
  \end{cases} \label{Scons}
\end{equation}
\begin{table}[!h]
\centering
\caption{Simulated profile data set in phase $I$ are provided here to calculate $UCL_{ME}$ and $UCL_{LR}$ with $\overline{x}=0.125$.}\label{Sdata}
\begin{tabular}{|c|c|c|c|c|c|c|} \hline
$y$	& $x=0.05$	& $x=0.1$	& $x=0.15$	& $x=0.2$ & $\overline{y}$ & $ \overline{xy} $ \\ \hline
$1$	& $0.135$	& $1.434$	& $1.228$	& $2.133$ & $1.2325$ & $0.1902375$ \\
$2$	& $0.955$	& $1.143$	& $1.493$	& $2.058$ & $1.41225$ & $0.1994$ \\
$3$	& $0.267$	& $1.122$	& $1.35$	& $1.948$ & $1.17175$ & $0.1794125$ \\
$4$	& $0.179$	& $0.915$	& $1.154$	& $2.435$ & $1.17075$ & $0.1901375$ \\ \hline
\end{tabular}
\end{table}\\
Then, the Lagrange functions have to be written for each sample $j=1,\ldots ,4$:
\begin{eqnarray}\label{L}
L(f,\lambda_0,~\lambda_1,~\lambda_2,~\lambda_3)&=&-\int\int_{\mathbb{S}(X,Y)} \log f_{X,Y}(x,y) ~ dF_{X,Y}(x,y)-\lambda_0 \{ \int\int_{\mathbb{S}(X,Y)} dF_{X,Y}(x,y)-1\}   \nonumber\\
&-& \lambda_1 \{ \int\int_{\mathbb{S}(X,Y)} x~ dF_{X,Y}(x,y)-\overline{x}\}-\lambda_2 \{ \int\int_{\mathbb{S}(X,Y)} y ~dF_{X,Y}(x,y)-\overline{y}_j\} \nonumber\\
&-& \lambda_3 \{ \int\int_{\mathbb{S}(X,Y)} x~y~ dF_{X,Y}(x,y)-\overline{xy}_j\}. \nonumber
 \end{eqnarray}
Now, after deriving and placing zero, like the steps in Section \ref{ME}, we get the following functions, which is a similar equation to \eqref{MEf} as well:
\begin{equation}\label{SMEf}
 f_{X,Y_j}(x,y) = \exp (-1-\lambda_0-x\lambda_1-y\lambda_2-xy\lambda_3),~~ (x,y)\in \mathbb{S}(X,Y),~~j=1,\ldots,4.
\end{equation}
In the next step, the conditions will be upgraded using function \eqref{SMEf} as following for $j=1,~\ldots,~4$:
\begin{equation}
 \begin{cases}
\int\int_{\mathbb{S}(X,Y)} \exp (-1-\lambda_0-x\lambda_1-y\lambda_2-xy\lambda_3)dxdy=1, \\
\int\int_{\mathbb{S}(X,Y)} x \exp (-1-\lambda_0-x\lambda_1-y\lambda_2-xy\lambda_3)dxdy=0.125, \\
\int\int_{\mathbb{S}(X,Y)} y_j \exp (-1-\lambda_0-x\lambda_1-y\lambda_2-xy\lambda_3)dxdy=\overline{y}_j, \\
\int\int_{\mathbb{S}(X,Y)} x y_j \exp (-1-\lambda_0-x\lambda_1-y\lambda_2-xy\lambda_3)dxdy=\overline{x y}_j. \\
  \end{cases} \label{SCxyf}
\end{equation}
So, the system of equations \eqref{SCxyf} contains four unknown parameters $\lambda_0,\ldots,\lambda_3$, and has to be solved concerning them. The number of samples is $4$, so system \eqref{SCxyf} has to be solved four times to find out the maximum entropy distributions of $4$ samples. All Lagrange coefficients are available in Table \ref{Slambda} as well. For instance, the maximum entropy distribution of the first sample is:
\begin{equation}\label{fs1}
 f_{X,Y_1}(x,y) = \exp (-1+3.070517-7.57113~x-0.9876531~y-0.4731329~x~y),~~ (x,y)\in \mathbb{S}(X,Y).
\end{equation}
\begin{table}[!h]
\centering
\caption{The coefficients of maximum entropy distributions are calculated via solving systems of equations containing $4$ unknown parameters $\lambda_0$, $\lambda_1$, $\lambda_2$, and $\lambda_3$.}\label{Slambda}
\begin{tabular}{|c|c|c|c|c|} \hline
Groups	& $\lambda_0$	& $\lambda_1$	& $\lambda_2$	& $\lambda_3$ \\ \hline
$1$	& $-3.070517$	& $7.57113$	& $0.9876531$	& $0.4731329$ \\
$2$	& $-2.840981$	& $7.714995$	& $0.7845862$	& $0.2433592$ \\
$3$	& $-3.114455$	& $7.584808$	& $1.0323813$	& $0.476777$ \\
$4$	& $-3.167204$	& $7.50769$	& $1.0872566$	& $0.6066953$ \\ \hline
\end{tabular}
\end{table}
The next stage is to get the profile coefficients via the maximum entropy method using the obtained distributions for samples. The formulas \eqref{BhadM} and \eqref{AhadM} are used to figure out $ \underset{\sim}{m_1},~\ldots,~\underset{\sim}{m_k} $. For example, the procedure to get $\underset{\sim}{m_1}$ is:
\begin{equation*}
  EX=\int\int_{\mathbb{S}(X,Y)} x~ f_{X,Y_1}(x,y)dxdy=0.125,
\end{equation*}
\begin{equation*}
 E[(X-EX)^2] = \int\int_{\mathbb{S}(X,Y)} (x-EX)^2 f_{X,Y_1}(x,y)dxdy=0.01573245,
 \end{equation*}
\begin{equation*}
  EY_1=\int\int_{\mathbb{S}(X,Y)} y~ f_{X,Y_1}(x,y)dxdy=0.9547399,
\end{equation*}
\begin{equation*}
  \widehat{b}_{1-ME}=\frac{1}{E[(X-EX)^2]}\int\int_{\mathbb{S}(X,Y)} (x-EX)(y-EY_1) ~f_{X,Y_1}(x,y)dxdy=-0.3843313,
\end{equation*}
and
\begin{equation*}
  \widehat{a}_{1-ME}= EY_1-\widehat{b}_{1-ME}~EX=1.0027813,
\end{equation*}
where $f_{X,Y_1}(\cdot,\cdot)$ is defined in equation \eqref{fs1}. All steps are done for samples, and the results are in Table \ref{sT2ME}. $T^2 _{ME}$s are based on the formula \eqref{T2fM}. The upper control limit $UCL_{ME}$ is the quantile $95\%$ of the $T^2_{ME}$s, which is $1.963768$.
\begin{table}[!h]
\centering
\caption{Estimated profile coefficients are calculated via maximum entropy method. The last column is contained their corresponding $T^2$-Hotelling values.}\label{sT2ME}
\begin{tabular}{|c|c|c|c|} \hline
Groups	& $\widehat{a}_{ME}$	& $\widehat{b}_{ME}$	& $T^2_{ME}$ \\ \hline
$1$	& $1.0027813$	& $-0.3843313$	& $0.8160215$ \\
$2$	& $1.2503569$	& $-0.3165483$	& $1.9671948$ \\
$3$	& $0.9607011$	& $-0.3586005$	& $1.9443471$ \\
$4$	& $0.9114218$	& $-0.3984706$	& $0.9042891$ \\ \hline
\end{tabular}
\end{table}
\begin{table}[!h]
\centering
\caption{Estimated profile coefficients are calculated via linear regression method. The last column is contained their corresponding $T^2$-Hotelling values.}\label{sT2LR}
\begin{tabular}{|c|c|c|c|} \hline
Groups	& $\widehat{a}_{LR}$	& $\widehat{b}_{LR}$	& $T^2_{LR}$ \\ \hline
$1$	& $-0.2145$	& $11.576$	& $0.1067833$ \\
$2$	& $0.4975$	& $7.318$	& $2.0360719$ \\
$3$	& $-0.146$	& $10.542$	& $1.4706997$ \\
$4$	& $-0.581$	& $14.014$	& $1.8363376$ \\ \hline
\end{tabular}
\end{table}\\
The steps are so simple for the linear regression as a classical statistical method. The result of this procedure is in Table \ref{sT2LR}. Its upper control $UCL_{LR}$ is the quantile $95\%$ of the $T^2_{LR}$s and is $2.006112$. The upper control limit based on Fisher distribution is $71.25$, so far from logic. The same steps have to be done for data in phase $II$ to find the maximum entropy distributions and their corresponding profile coefficients. The difference between phase $I$ and phase $II$ calculation is related to get the values of $T^2$-Hotelling statistics. In this case, the formulations \eqref{T2fM} and \eqref{T2fL} are applied with $S_m$, $\underset{\sim}{\overline{m}}$, $S_l$, and $\underset{\sim}{\overline{l}}$ from the phase $I$ data set. More information is available in Montgomery $(2007)$.\\
Let's show the performance of the methods by calculating $ARL$ for the simulation data set. To start the main simulation study, first of all, we simulate $ 100 $ samples with size $ n=5 $ from model $ Y=2+3X+\varepsilon $, where $ \varepsilon $ has a normal distribution of mean $ 0 $ and variance $ 0.1 $. The fixed observation vector of $ X $ for all $100$ samples is $ (2,~2.2,~2.4,~2.1,~2.7) $. For these datasets, we calculate upper control limits of Fisher distribution, the $ 95 $ present quantile in maximum entropy, and linear regression respectively:
\begin{displaymath}
\left\{\begin{array}{ll}
UCL_{F}=6.303865, \\
UCL_{ME}=5.591411, \\
UCL_{LR}=5.80042. \\
\end{array}\right.
\end{displaymath}
Then, $ 1000 $ sets of the correct model are simulated again to calculate $ ARL_0 $. These are $ARL_0$ for the maximum entropy technique:
\begin{displaymath}
ARL_{0}=\left\{\begin{array}{ll}
\frac{1}{0.059}=16.949\simeq 17, & UCL=UCL_F , \\
\\
\frac{1}{0.08}=12.5\simeq 13, & UCL=UCL_{ME}, \\
\end{array}\right.
\end{displaymath}
and the related $ARL_0$ for the linear regression method are:
\begin{displaymath}
ARL_{0}=\left\{\begin{array}{ll}
\frac{1}{0.06}=16.666\simeq 17, & UCL=UCL_F, \\
\\
\frac{1}{0.071}=14.084\simeq 15, & UCL=UCL_{LR}. \\
\end{array}\right.
\end{displaymath}
They are the number of previous samples to observe one point out of the $ UCL $ limit that means the false alarm when the process is in an appropriate situation because we simulate all datasets from the true model. This means that the process is in statistical control, so the larger value of $ARL_0$, the better choice of control limits. Therefore false alarm occurs later and less. It can be concluded from $ ARL_0 $s that when the $ UCL $ is fewer and more sensitive, the first type of error is increased. On the other, it is reduced when the corresponding $ UCL $ is less sensitive. Now we are keen on simulate datasets in shifted models and testing these two methods to detect the changes of the process means. Here we provide three shifted models containing intercepts, slopes, and a mixture of intercepts and slopes shifts as follows:
\begin{displaymath}
\left\{\begin{array}{ll}
I:~ Y=(2+s)+3X+\varepsilon, \\
II:~ Y=2+(3+s)X+\varepsilon, \\
III:~ Y=(2+s)+(3+s)X+\varepsilon, \\
\end{array}\right.
\end{displaymath}
where shift $ s $ is started from $ 0.01 $ till $ 0.32 $ and the steps are $ 0.01 $:
\begin{equation*}
  s=\{0.01,~0.02,~0.03,~\ldots,~~0.3,~0.31,~0.32\}.
\end{equation*}
The second type of error $ \beta $ is plotted for three shifted models in Figures \ref{1}, \ref{2}, and \ref{3}. The horizontal axis is determined for different amounts of shift $ s $, and the vertical axis is defined for $\beta$ related to several shifts. In each plot, we present four curves. Two of them are for the maximum entropy method with different upper control limits, and the other two curves are for the linear regression technique specified by the legends on the right top of each plot. We can derive from all plots that the more shifts, the more sensitive control limits for detecting changes in a procedure. In Figure \ref{1}, the best method is the maximum entropy principle while using $UCL_{ME}$ quantile, and the worse model to detect shifts is the linear regression method using the upper Fisher control limit $UCL_F$, and their discrepancy is obvious on the figure. In Figures \ref{2} and \ref{3}, this discrepancy is becoming lesser, because slope sifts effect more on process data, and in the third plot, there are intercept and slope shifts together. Eventually, we can conclude based on the curves that when shifts increase, the probabilities of false alarms decrease. Furthermore, we provide $ ARL_1 $s for all three models in Tables \ref{ti}, \ref{tc}, and \ref{ts} which are based on several shifts in intercepts, slopes, and a mixed model of intercepts and slope. The same conclusion of plots can drown from tables too. As it can be seen in the figures and the tables, these two detections ways are proper for smooth shifts, but the way based on maximum entropy discovers much sooner than the linear regression method. These ways can be applied in some manufacturing processes which should not have small shifts and always should be insensitive desirable situations like pharmaceutical products or expensive procedures, potential changes should be detected earlier to avoid wasting wealth.
\begin{table}[!h]
\centering
\caption{$ ARL_1 $s are provided for intercept Shifts where the columns $ LR_F $ and $ LR_q $ are related to linear regression method using $UCL_F$ and $UCL_{LR}$, and columns $ ME_F $ and $ ME_q $ are for maximum entropy principle with two upper control limits $UCL_F$ and $UCL_{ME}$ respectively.}\label{ti}
\begin{tabular}{|c|c|c|c|c|}
\hline
\multirow{2}{*}{$Shifts$} &\multicolumn{2}{|c|}{$LR$ method} & \multicolumn{2}{|c|}{$ME$ method} \\ \cline{2-5}
 & $ UCL_F $ & $ UCL_{LR} $ & $ UCL_F $ & $ UCL_{ME} $ \\
\hline
  $ 0.01 $ & $ 19.607 $ & $ 15.625 $ & $ 17.543 $ & $ 12.500 $ \\
  $ 0.02 $ & $ 20.408 $ & $ 18.518 $ & $ 14.925 $ & $ 11.363 $ \\
  $ 0.03 $ & $ 14.492 $ & $ 12.048 $ & $ 10.101 $ & $ 8.620 $ \\
  $ 0.04 $ & $ 13.698 $ & $ 11.494 $ & $ 10.989 $ & $ 8.333 $ \\
  $ 0.05 $ & $ 12.345 $ & $ 10.309 $ & $ 9.433 $ & $ 7.299 $ \\
  $ 0.06 $ & $ 9.259 $ & $ 7.462 $ & $ 7.299 $ & $ 6.060 $ \\
  $ 0.07 $ & $ 9.174 $ & $ 7.936 $ & $ 6.896 $ & $ 5.847 $ \\
  $ 0.08 $ & $ 7.633 $ & $ 6.329 $ & $ 6.060 $ & $ 4.784 $ \\
  $ 0.09 $ & $ 6.622 $ & $ 5.494 $ & $ 5.000 $ & $ 3.861 $ \\
  $ 0.10 $ & $ 5.747 $ & $ 4.807 $ & $ 4.424 $ & $ 3.484 $ \\
  $ 0.11 $ & $ 4.950 $ & $ 4.219 $ & $ 3.731 $ & $ 3.134 $ \\
  $ 0.12 $ & $ 3.472 $ & $ 3.030 $ & $ 2.890 $ & $ 2.433 $ \\
  $ 0.13 $ & $ 2.941 $ & $ 2.583 $ & $ 2.525 $ & $ 2.114 $ \\
  $ 0.14 $ & $ 2.500 $ & $ 2.197 $ & $ 2.100 $ & $ 1.808 $ \\
  $ 0.15 $ & $ 2.016 $ & $ 1.821 $ & $ 1.745 $ & $ 1.557 $ \\
  $ 0.16 $ & $ 1.776 $ & $ 1.647 $ & $ 1.605 $ & $ 1.449 $ \\
  $ 0.17 $ & $ 1.594 $ & $ 1.455 $ & $ 1.445 $ & $ 1.310 $ \\
  $ 0.18 $ & $ 1.418 $ & $ 1.342 $ & $ 1.295 $ & $ 1.206 $ \\
  $ 0.19 $ & $ 1.270 $ & $ 1.210 $ & $ 1.183 $ & $ 1.112 $ \\
  $ 0.20 $ & $ 1.221 $ & $ 1.169 $ & $ 1.165 $ & $ 1.119 $ \\
  $ 0.21 $ & $ 1.161 $ & $ 1.140 $ & $ 1.121 $ & $ 1.075 $ \\
  $ 0.22 $ & $ 1.094 $ & $ 1.068 $ & $ 1.070 $ & $ 1.040 $ \\
  $ 0.23 $ & $ 1.066 $ & $ 1.054 $ & $ 1.049 $ & $ 1.033 $ \\
  $ 0.24 $ & $ 1.038 $ & $ 1.024 $ & $ 1.023 $ & $ 1.015 $ \\
  $ 0.25 $ & $ 1.017 $ & $ 1.011 $ & $ 1.007 $ & $ 1.004 $ \\
  $ 0.26 $ & $ 1.015 $ & $ 1.010 $ & $ 1.010 $ & $ 1.003 $ \\
  $ 0.27 $ & $ 1.007 $ & $ 1.005 $ & $ 1.004 $ & $ 1.002 $ \\
  $ 0.28 $ & $ 1.002 $ & $ 1.001 $ & $ 1.001 $ & $ 1.001 $ \\
  $ 0.29 $ & $ 1.001 $ & $ 1.000 $ & $ 1.000 $ & $ 1.000 $ \\
  $ 0.30 $ & $ 1.001 $ & $ 1.000 $ & $ 1.000 $ & $ 1.000 $ \\
  $ 0.31 $ & $ 1.000 $ & $ 1.000 $ & $ 1.000 $ & $ 1.000 $ \\
  $ 0.32 $ & $ 1.000 $ & $ 1.000 $ & $ 1.000 $ & $ 1.000 $\\
\hline
\end{tabular}
\end{table}
\begin{table}[!h]
\centering
\caption{$ ARL_1 $s are presented for slope shifts in several ways using different $UCL$s which are describe before.}\label{tc}
\begin{tabular}{|c|c|c|c|c|}
\hline
\multirow{2}{*}{$Shifts$} &\multicolumn{2}{|c|}{$LR$ method} & \multicolumn{2}{|c|}{$ME$ method} \\ \cline{2-5}
 & $ UCL_F $ & $ UCL_{LR} $ & $ UCL_F $ & $ UCL_{ME} $ \\
\hline
  $ 0.01 $ & $ 21.276 $ & $ 17.241 $ & $ 13.888 $ & $ 10.526 $ \\
  $ 0.02 $ & $ 11.235 $ & $ 9.433 $ & $ 9.174 $ & $ 7.299 $ \\
  $ 0.03 $ & $ 6.756 $ & $ 5.882 $ & $ 5.952 $ & $ 4.651 $ \\
  $ 0.04 $ & $ 5.025 $ & $ 4.464 $ & $ 4.901 $ & $ 3.676 $ \\
  $ 0.05 $ & $ 3.448 $ & $ 3.048 $ & $ 3.012 $ & $ 2.493 $ \\
  $ 0.06 $ & $ 2.415 $ & $ 2.188 $ & $ 2.192 $ & $ 1.886 $ \\
  $ 0.07 $ & $ 1.600 $ & $ 1.464 $ & $ 1.485 $ & $ 1.353 $ \\
  $ 0.08 $ & $ 1.293 $ & $ 1.256 $ & $ 1.264 $ & $ 1.199 $ \\
  $ 0.09 $ & $ 1.102 $ & $ 1.078 $ & $ 1.081 $ & $ 1.053 $ \\
  $ 0.10 $ & $ 1.049 $ & $ 1.035 $ & $ 1.033 $ & $ 1.015 $ \\
  $ 0.11 $ & $ 1.020 $ & $ 1.010 $ & $ 1.012 $ & $ 1.008 $ \\
  $ 0.12 $ & $ 1.002 $ & $ 1.002 $ & $ 1.001 $ & $ 1.000 $ \\
  $ 0.13 $ & $ 1.000 $ & $ 1.000 $ & $ 1.000 $ & $ 1.000 $ \\
  $ 0.14 $ & $ 1.000 $ & $ 1.000 $ & $ 1.000 $ & $ 1.000 $ \\
\hline
\end{tabular}
\end{table}
\begin{table}[!h]
\centering
\caption{$ ARL_1 $s are exhibited for the mixed model of intercept and slope shifts in several columns which are explained in the  caption of Table \ref{ti}.}\label{ts}
\begin{tabular}{|c|c|c|c|c|}
\hline
\multirow{2}{*}{$Shifts$} &\multicolumn{2}{|c|}{$LR$ method} & \multicolumn{2}{|c|}{$ME$ method} \\ \cline{2-5}
 & $ UCL_F $ & $ UCL_{LR} $ & $ UCL_F $ & $ UCL_{ME} $ \\
\hline
  $ 0.01 $ & $ 15.384 $ & $ 12.658 $ & $ 12.048 $ & $ 10.101 $ \\
  $ 0.02 $ & $ 10.416 $ & $ 8.620 $ & $ 7.751 $ & $ 6.211 $ \\
  $ 0.03 $ & $ 4.716 $ & $ 4.132 $ & $ 3.937 $ & $ 3.174 $ \\
  $ 0.04 $ & $ 2.439 $ & $ 2.222 $ & $ 2.192 $ & $ 1.912 $ \\
  $ 0.05 $ & $ 1.557 $ & $ 1.474 $ & $ 1.445 $ & $ 1.344 $ \\
  $ 0.06 $ & $ 1.221 $ & $ 1.149 $ & $ 1.160 $ & $ 1.111 $ \\
  $ 0.07 $ & $ 1.044 $ & $ 1.029 $ & $ 1.031 $ & $ 1.018 $ \\
  $ 0.08 $ & $ 1.009 $ & $ 1.006 $ & $ 1.008 $ & $ 1.003 $ \\
  $ 0.09 $ & $ 1.002 $ & $ 1.001 $ & $ 1.001 $ & $ 1.000 $ \\
  $ 0.10 $ & $ 1.000 $ & $ 1.000 $ & $ 1.000 $ & $ 1.000 $ \\
  $ 0.11 $ & $ 1.000 $ & $ 1.000 $ & $ 1.000 $ & $ 1.000 $ \\
\hline
\end{tabular}
\end{table}

\section{\bf Real dataset on semiconductor production process}\label{6}
\begin{figure}[!h]
\begin{center}
\centering \vspace*{-.5cm}
\includegraphics[width=160mm]{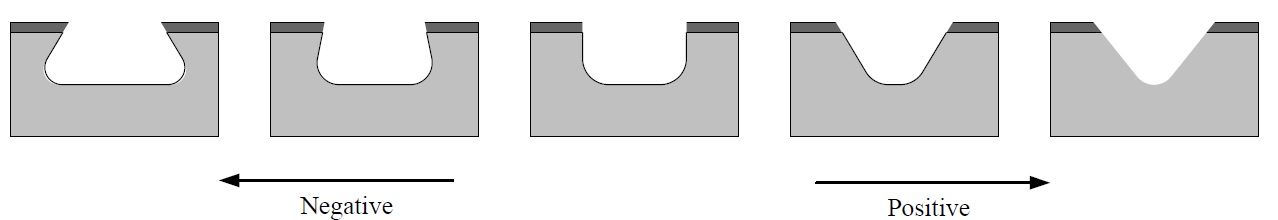}
\caption{Illustrations of various etching profiles form of a $(DRIE)$ process. (Adapted from Zou et al. $(2007)$.)}\label{Zou}
\end{center}
\end{figure}

\begin{figure}[!h]
\begin{center}
\centering \vspace*{-.5cm}
\includegraphics[width=90mm]{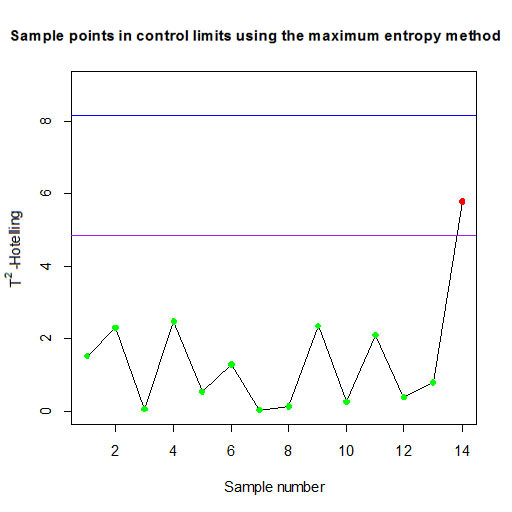}
\caption{The control limits for the second phase dataset using the maximum entropy method are drawn here. The blue and purple lines are $UCL_F$ and $UCL_{ME}$ respectively. All points of samples are in control with the control limit $UCL_F$ which is based on Fisher distribution, but the red point related to the last sample is out of control based on $ UCL_{ME}$.}\label{PlotP1}
\end{center}
\end{figure}

In the manufacturing process, detecting all unexpected shifts is very important. There are many different ways for different situations to deal with out of control processes. The technicians have to choose a controlling method which detects them as soon as possible. There are some restrictions on the result products of processes that unwanted changes are forbidden even for some small shift. Many classical methods based on the mean control charts are disabled to detect a soft shift. In this paper, we explain two methods of detecting soft changes, the maximum entropy manner, and the classical linear regression. We compared them via simulation study in section \ref{5}, but here a real data example is represented. The initial profile dataset is from Zou et al. $(2007)$, whose process is on deep reactive ion etching process $(DRIE)$ from semiconductor manufacturing that contains two phases $I$ and $II$. They mentioned in their paper about the source of data that "in the $(DRIE)$ process, one of the most important quality characteristics is the profile of a trench that may significantly impact the downstream operations (May et al. $(1991)$). The desired profile is the one with a smooth and vertical sidewall, as indicated in the center sample of Figure \ref{Zou}". The first dataset is included $18$ samples with size $n=11$, and the observed vector of independent variable is $ (-2.5,-2,-1.5,-1,-0.5,0,0.5,1,1.5,2,2.5) $. The second set that is for phase $II$ has consisted of $14$ samples of the same size as the first set. \\
\\
We aim to check the second phase of the process via suitable statistical control limits to see the possible defective products. By the way, in Zou et al. $(2007)$, they noted that the last sample of Phase $II$ with the number $14$ is out of control. Thus, our control limits should detect these changes in the process. Therefore, we calculate the $UCL_F$ control limit based on Fisher distribution and compare it with our presented $UCL_{ME}$ and $UCL_{LR}$. It is worth mentioning again that the control limit $UCL_F$ is based on mean control, but other control limits are based on profile coefficients control limits. Hence, to find out the $UCL_{ME}$ limit, we have to compute the maximum entropy distributions of all $18$ samples. We provide the Lagrange coefficients of the maximum entropy principle according to constraints \eqref{constraints} in Table \ref{MeP1}. There are six coefficients. Based on what we said in section \ref{CalcuCO} and \ref{ME}, $\lambda_0$ guarantees the maximum entropy distribution to be valid. Then, profile coefficients of all $18$ maximum entropy distributions are calculated in Table \ref{CoMeP1} as well with their $T^2$-Hotelling values. The coefficients of the regression method are provided for this sample in Table \ref{CoLrP1}, and their $T^2$-Hotellings are in the last column on the right for each sample. Till here, we get some important information form the situation of the in-control process. The next step is to specify beneficial control limits to use them for monitoring the rest of the process data. Thus, we obtain:
\begin{displaymath}
\left\{\begin{array}{ll}
UCL_{F}=8.150644, \\
UCL_{ME}=4.857291, \\
UCL_{LR}=4.857291. \\
\end{array}\right.
\end{displaymath}
In the subsequent stage, we have to gather some information from phase $II$ data as before. We start it by calculation of maximum entropy Lagrange coefficients of the $14$ samples shown in Table \ref{MeP2}. Then, the profile coefficients and $T^2$-Hotellings are in Table \ref{CoMeP2}. In Table \ref{CoLrP2}, the profile coefficients are offered with their values of Hotelling statistics. The same conclusion of simulation results can be driven here. We know from Zou et al. $(2007)$ that the last sample of phase $II$ is out of control, and by using the two presented methods, the same consequence is gotten. The last one is out of control, as can be seen in Figure \ref{PlotP1}. The interesting part of the result is that the traditional method based on Fisher is unable to detect this shift.
\begin{table}[!h]
\centering
\caption{The Lagrange coefficients of the maximum entropy distribution based on the in-control situation are here.}\label{MeP1}
\begin{tabular}{|c|c|c|c|c|c|c|}
\hline
Sample number & $\lambda_0$ & $\lambda_1$ & $\lambda_2$ & $\lambda_3$ & $\lambda_4$ & $\lambda_5$ \\ \hline
$1$ & $2.148452$ & $0.086787$ & $-0.73435$ & $0.203605$ & $0.258242$ & $-0.06104$ \\ \hline
$2$ & $2.153768$ & $-0.01368$ & $-0.75237$ & $0.200085$ & $0.263399$ & $0.009578$ \\ \hline
$3$ & $2.150188$ & $-0.07209$ & $-0.72613$ & $0.202509$ & $0.254864$ & $0.050602$ \\ \hline
$4$ & $2.274768$ & $-0.04345$ & $-0.81836$ & $0.200726$ & $0.258234$ & $0.02742$ \\ \hline
$5$ & $2.126714$ & $0.021106$ & $-0.84734$ & $0.200189$ & $0.306399$ & $-0.01526$ \\ \hline
$6$ & $2.293237$ & $0.023812$ & $-0.80842$ & $0.200215$ & $0.250636$ & $-0.01476$ \\ \hline
$7$ & $2.082117$ & $0.064123$ & $-0.56338$ & $0.202801$ & $0.216233$ & $-0.04922$ \\ \hline
$8$ & $2.340003$ & $0.097769$ & $-0.9068$ & $0.203203$ & $0.275546$ & $-0.05942$ \\ \hline
$9$ & $2.265429$ & $-0.08991$ & $-0.70341$ & $0.203606$ & $0.220693$ & $0.056417$ \\ \hline
$10$ & $2.202199$ & $-0.01261$ & $-0.7537$ & $0.200068$ & $0.251997$ & $0.00843$ \\ \hline
$11$ & $2.195643$ & $-0.16325$ & $-0.85185$ & $0.210633$ & $0.289564$ & $0.110982$ \\ \hline
$12$ & $2.298673$ & $0.085613$ & $-0.79945$ & $0.202822$ & $0.246324$ & $-0.05276$ \\ \hline
$13$ & $2.397235$ & $0.052969$ & $-0.8115$ & $0.200982$ & $0.230539$ & $-0.0301$ \\ \hline
$14$ & $2.307133$ & $-0.00526$ & $-0.99678$ & $0.200009$ & $0.317079$ & $0.003344$ \\ \hline
$15$ & $2.247514$ & $-0.0266$ & $-0.56266$ & $0.200402$ & $0.179815$ & $0.017001$ \\ \hline
$16$ & $2.378963$ & $-0.04418$ & $-0.73862$ & $0.200752$ & $0.210053$ & $0.02513$ \\ \hline
$17$ & $2.56554$ & $-0.07458$ & $-1.13947$ & $0.201342$ & $0.313354$ & $0.041021$ \\ \hline
$18$ & $2.103819$ & $0.001211$ & $-0.60548$ & $0.2$ & $0.224705$ & $-0.0009$ \\ \hline
\end{tabular}
\end{table}
\begin{table}[!h]
\centering
\caption{The maximum entropy coefficients for the $18$ samples of in-control data gathering with their corresponding $T^2$-Hotellings are as below.}\label{CoMeP1}
\begin{tabular}{|c|c|c|c|}
\hline
Sample number & $\widehat{a}_{ME}$ & $\widehat{b}_{ME}$ & $T^2_{ME}$ \\ \hline
$1$ & $1.421818$ & $0.118182$ & $2.863782$ \\ \hline
$2$ & $1.428182$ & $-0.01818$ & $0.700087$ \\ \hline
$3$ & $1.424545$ & $-0.09927$ & $1.985242$ \\ \hline
$4$ & $1.584545$ & $-0.05309$ & $0.376158$ \\ \hline
$5$ & $1.382727$ & $0.024909$ & $1.416012$ \\ \hline
$6$ & $1.612727$ & $0.029455$ & $0.422706$ \\ \hline
$7$ & $1.302727$ & $0.113818$ & $4.802159$ \\ \hline
$8$ & $1.645455$ & $0.107818$ & $2.398796$ \\ \hline
$9$ & $1.593636$ & $-0.12782$ & $2.165594$ \\ \hline
$10$ & $1.495455$ & $-0.01673$ & $0.137896$ \\ \hline
$11$ & $1.470909$ & $-0.19164$ & $5.169704$ \\ \hline
$12$ & $1.622727$ & $0.107091$ & $2.161098$ \\ \hline
$13$ & $1.76$ & $0.065273$ & $3.098238$ \\ \hline
$14$ & $1.571818$ & $-0.00527$ & $0.0367$ \\ \hline
$15$ & $1.564545$ & $-0.04727$ & $0.25041$ \\ \hline
$16$ & $1.758182$ & $-0.05982$ & $2.605809$ \\ \hline
$17$ & $1.818182$ & $-0.06545$ & $4.121102$ \\ \hline
$18$ & $1.347273$ & $0.002$ & $1.936388$ \\ \hline
\end{tabular}
\end{table}
\begin{table}[!h]
\centering
\caption{The coefficients of the linear regression method whose $T^2$-Hotelling values are on the right column are provided for the first sample.}\label{CoLrP1}
\begin{tabular}{|c|c|c|c|}
\hline
Sample number & $\widehat{a}_{LR}$ & $\widehat{b}_{LR}$ & $T^2_{LR}$ \\ \hline
$1$ & $1.421818$ & $0.118182$ & $2.863775$ \\ \hline
$2$ & $1.428182$ & $-0.01818$ & $0.700087$ \\ \hline
$3$ & $1.424545$ & $-0.09927$ & $1.98524$ \\ \hline
$4$ & $1.584545$ & $-0.05309$ & $0.376158$ \\ \hline
$5$ & $1.382727$ & $0.024909$ & $1.416012$ \\ \hline
$6$ & $1.612727$ & $0.029455$ & $0.422706$ \\ \hline
$7$ & $1.302727$ & $0.113818$ & $4.802158$ \\ \hline
$8$ & $1.645455$ & $0.107818$ & $2.398797$ \\ \hline
$9$ & $1.593636$ & $-0.12782$ & $2.165596$ \\ \hline
$10$ & $1.495455$ & $-0.01673$ & $0.137896$ \\ \hline
$11$ & $1.470909$ & $-0.19164$ & $5.169706$ \\ \hline
$12$ & $1.622727$ & $0.107091$ & $2.161099$ \\ \hline
$13$ & $1.76$ & $0.065273$ & $3.098237$ \\ \hline
$14$ & $1.571818$ & $-0.00527$ & $0.0367$ \\ \hline
$15$ & $1.564545$ & $-0.04727$ & $0.25041$ \\ \hline
$16$ & $1.758182$ & $-0.05982$ & $2.605808$ \\ \hline
$17$ & $1.818182$ & $-0.06545$ & $4.121102$ \\ \hline
$18$ & $1.347273$ & $0.002$ & $1.936388$ \\ \hline
\end{tabular}
\end{table}
\begin{table}[!h]
\centering
\caption{The Lagrange coefficients of the maximum entropy for each sample of phase $II$ are calculated.}\label{MeP2}
\begin{tabular}{|c|c|c|c|c|c|c|}
\hline
Sample number & $\lambda_0$ & $\lambda_1$ & $\lambda_2$ & $\lambda_3$ & $\lambda_4$ & $\lambda_5$ \\ \hline
$1$ & $2.350736$ & $0.094391$ & $-1.05091$ & $0.202642$ & $0.32748$ & $-0.05883$ \\ \hline
$2$ & $2.2661$ & $-0.10878$ & $-0.81399$ & $0.204617$ & $0.258632$ & $0.069125$ \\ \hline
$3$ & $2.237455$ & $-0.01834$ & $-0.72038$ & $0.200149$ & $0.232379$ & $0.01183$ \\ \hline
$4$ & $2.217579$ & $0.100131$ & $-0.7993$ & $0.204152$ & $0.264668$ & $-0.06631$ \\ \hline
$5$ & $2.227532$ & $0.038925$ & $-0.70193$ & $0.200701$ & $0.228301$ & $-0.02532$ \\ \hline
$6$ & $2.173455$ & $0.047915$ & $-0.63502$ & $0.201241$ & $0.218015$ & $-0.0329$ \\ \hline
$7$ & $2.247015$ & $-0.00031$ & $-0.56221$ & $0.200001$ & $0.179777$ & $0.000196$ \\ \hline
$8$ & $2.211425$ & $-0.02008$ & $-0.63823$ & $0.200209$ & $0.210699$ & $0.013255$ \\ \hline
$9$ & $2.273512$ & $-0.07685$ & $-0.5887$ & $0.203114$ & $0.182724$ & $0.047708$ \\ \hline
$10$ & $2.277067$ & $0.011272$ & $-0.73804$ & $0.200052$ & $0.229853$ & $-0.00702$ \\ \hline
$11$ & $2.140158$ & $0.05129$ & $-0.61059$ & $0.201535$ & $0.217643$ & $-0.03656$ \\ \hline
$12$ & $2.267622$ & $0.018718$ & $-0.56256$ & $0.200195$ & $0.176$ & $-0.01171$ \\ \hline
$13$ & $2.32175$ & $0.019312$ & $-0.7868$ & $0.200142$ & $0.237118$ & $-0.01164$ \\ \hline
$14$ & $2.153608$ & $-0.08341$ & $-0.46953$ & $0.205404$ & $0.171248$ & $0.06084$ \\ \hline
\end{tabular}
\end{table}
\begin{table}[!h]
\centering
\caption{The profile coefficients of the maximum entropy manner for the sample in phase $II$ with their relative $T^2$-Hotelling values are gotten as below.}\label{CoMeP2}
\begin{tabular}{|c|c|c|c|}
\hline
Sample number & $\widehat{a}_{ME}$ & $\widehat{b}_{ME}$ & $T^2_{ME}$ \\ \hline
$1$ & $1.604545$ & $0.089818$ & $1.511138$ \\ \hline
$2$ & $1.573636$ & $-0.13364$ & $2.303208$ \\ \hline
$3$ & $1.550002$ & $-0.02545$ & $0.051808$ \\ \hline
$4$ & $1.51$ & $0.125273$ & $2.482413$ \\ \hline
$5$ & $1.537273$ & $0.055455$ & $0.54043$ \\ \hline
$6$ & $1.456364$ & $0.075455$ & $1.297488$ \\ \hline
$7$ & $1.563636$ & $-0.00055$ & $0.023144$ \\ \hline
$8$ & $1.514545$ & $-0.03145$ & $0.137063$ \\ \hline
$9$ & $1.610909$ & $-0.13055$ & $2.347447$ \\ \hline
$10$ & $1.605455$ & $0.015273$ & $0.255684$ \\ \hline
$11$ & $1.402727$ & $0.084$ & $2.090796$ \\ \hline
$12$ & $1.598182$ & $0.033273$ & $0.374238$ \\ \hline
$13$ & $1.659091$ & $0.024545$ & $0.801125$ \\ \hline
$14$ & $1.370909$ & $-0.17764$ & $\textbf{5.773461}$ \\ \hline
\end{tabular}
\end{table}
\begin{table}[!h]
\centering
\caption{The profile coefficients of the linear regression method for the sample in phase $II$ and their $T^2$-Hotellings are gathered here.}\label{CoLrP2}
\begin{tabular}{|c|c|c|c|}
\hline
Sample number & $\widehat{a}_{LR}$ & $\widehat{b}_{LR}$ & $T^2_{LR}$ \\ \hline
$1$ & $1.604545$ & $0.089818$ & $1.511138$ \\ \hline
$2$ & $1.573636$ & $-0.13364$ & $2.303204$ \\ \hline
$3$ & $1.55$ & $-0.02545$ & $0.051807$ \\ \hline
$4$ & $1.51$ & $0.125273$ & $2.48241$ \\ \hline
$5$ & $1.537273$ & $0.055455$ & $0.54043$ \\ \hline
$6$ & $1.456364$ & $0.075455$ & $1.297487$ \\ \hline
$7$ & $1.563636$ & $-0.00055$ & $0.023144$ \\ \hline
$8$ & $1.514545$ & $-0.03145$ & $0.137063$ \\ \hline
$9$ & $1.610909$ & $-0.13055$ & $2.347444$ \\ \hline
$10$ & $1.605455$ & $0.015273$ & $0.255684$ \\ \hline
$11$ & $1.402727$ & $0.084$ & $2.090795$ \\ \hline
$12$ & $1.598182$ & $0.033273$ & $0.374238$ \\ \hline
$13$ & $1.659091$ & $0.024545$ & $0.801124$ \\ \hline
$14$ & $1.370909$ & $-0.17764$ & $\textbf{5.773456}$ \\ \hline
\end{tabular}
\end{table}\clearpage

\section{Pharmacy real profile data}\label{7}
As we know, the maximum entropy technique to estimate profile coefficients can be applied in pharmacy real data sets as well. In this paper, we explain two different ways to estimate profile coefficients. Then, we illustrate a simulation study and real data example on the semiconductor procedure. In these examples, we show the ability of the maximum entropy and linear regression to detect faults in production processes by defining some statistical control limits. In this section, we are keen on surveying another applied aspect of the maximum entropy method in manufacturing processes. The superiority of the maximum entropy principle is exhibited, while the regression method is unable to act well here. To do this, we choose a pharmaceutical industry profile dataset adapted from Shah et al. $(1998)$. In the pharmaceutical industry, there are many processes needed to be in statistical control. For instance, in the production of a particular tablet, many characteristics must be observed for the product to have a standard and desirable quality. Ma et al. $(2000)$ indicate that the U.S. Food and Drug Administration had to define many conductances of scale-up and post-approval changes for different types of shifted during production processes. In this regard, the statistical quality control charts have an influential position in pharmaceutical processes.\\
In Shah et al. $(1998)$, there are six groups of profile data containing the response variable in four columns as the cumulative dissolution of tablets at $30$, $60$, $90$, and $180$ minutes. The different values of time are the independent variable. The sample sizes have to be $12$ according to the guidance of "Dissolution Testing of Immediate Release Solid Oral Dosage Forms" Ma et al. $(2000)$. The first group is a pre-change as a dissolve reference in different minutes. The other five batches are from post-change processes. The beneficial aspect of this study was to detect the similarity or dissimilarity of batches from the reference group. So, we use the two explained methods here to realize if there are different or not. Thus, we calculate $T^2$-Hotelling statistics for these profiles via the maximum entropy and linear regression in Tables \ref{T2M} and \ref{T2L}. The upper control limits at the confidence level $0.9973$ are as below:
\begin{displaymath}
\left\{\begin{array}{ll}
UCL_{F}=26.97728, \\
UCL_{ME}=6.812622, \\
UCL_{LR}=5.883782. \\
\end{array}\right.
\end{displaymath}
\begin{table}[!h]
\centering
\caption{The $T^2 _{ME}$-Hotelling of the reference pre-change group and post-change batches $1$ to $5$ are calculated related to the tablet dissolution in $30$, $60$, $90$, and $180$ minutes. All numbers are out of control based on $UCL_{ME}$, but the stared numbers are in-control tablets according to $UCL_F$.}\label{T2M}
\begin{tabular}{|p{0.6in}|p{0.6in}|p{0.55in}|p{0.55in}|p{0.6in}|p{0.6in}|p{0.6in}|} \hline
Tablet numbers & Reference & Batch $1$ & Batch $2$ & Batch $3$ & Batch $4$ & Batch $5$ \\ \hline
$1$ & $0.909599$ & $2531250$ & $40212322$ & $24964618$ & $16.39547^*$ & $18160920$ \\ \hline
$2$ & $0.425575$ & $2528833$ & $40139281$ & $24937534$ & $19.67239^*$ & $18175083$ \\ \hline
$3$ & $0.941833$ & $2524719$ & $40201702$ & $24942289$ & $20.27384^*$ & $18143501$ \\ \hline
$4$ & $1.633346$ & $2524318$ & $40165632$ & $24924941$ & $13.65696^*$ & $18140501$ \\ \hline
$5$ & $6.866841$ & $2507804$ & $40231679$ & $24950972$ & $38.40114$ & $18161265$ \\ \hline
$6$ & $0.704702$ & $2503931$ & $40195131$ & $24925518$ & $20.16251^*$ & $18155740$ \\ \hline
$7$ & $1.089671$ & $2525108$ & $40156018$ & $24964433$ & $22.48306^*$ & $18159950$ \\ \hline
$8$ & $2.197987$ & $2524320$ & $40204586$ & $24949374$ & $17.02279^*$ & $18157899$ \\ \hline
$9$ & $0.374201$ & $2527543$ & $40223125$ & $24961179$ & $26.16568^*$ & $18151208$ \\ \hline
$10$ & $1.313367$ & $2529049$ & $40190721$ & $24955811$ & $23.06229^*$ & $18171547$ \\ \hline
$11$ & $5.041279$ & $2523157$ & $40213998$ & $24946082$ & $28.3081$ & $18163695$ \\ \hline
$12$ & $2.082611$ & $2512610$ & $40188450$ & $24942731$ & $37.34562$ & $18160788$ \\ \hline
\end{tabular}
\end{table}
\begin{table}[!h]
\centering
\caption{The $T^2 _{LR}$-Hotelling of the reference pre-change group and post-change batches $1$ to $5$ are calculated related to the tablet dissolution in $30$, $60$, $90$, and $180$ minutes. Bold numbers are in-control respect to $UCL_{LR}$, and the stared numbers are in-control tablets according to $UCL_F$}\label{T2L}
\begin{tabular}{|p{0.6in}|p{0.6in}|p{0.65in}|p{0.6in}|p{0.6in}|p{0.6in}|p{0.6in}|} \hline
Tablet numbers & Reference & Batch $1$ & Batch $2$ & Batch $3$ & Batch $4$ & Batch $5$ \\ \hline
$1$ & $0.753828$ & $\textbf{3.605572}^*$ & $9.745365^*$ & $22.38664^*$ & $26.5657^*$ & $30.8042$ \\ \hline
$2$ & $0.545729$ & $\textbf{1.257303}^*$ & $66.22903$ & $16.15048^*$ & $35.7986$ & $30.21156$ \\ \hline
$3$ & $1.065667$ & $\textbf{4.79904}^*$ & $10.03773^*$ & $16.77926^*$ & $46.38533$ & $55.70723$ \\ \hline
$4$ & $0.264546$ & $\textbf{3.286005}^*$ & $50.99222$ & $37.2213$ & $23.3263^*$ & $53.68816$ \\ \hline
$5$ & $5.944188$ & $63.69596$ & $7.94085^*$ & $20.63325^*$ & $43.57709$ & $25.90142^*$ \\ \hline
$6$ & $1.111635$ & $77.02181$ & $23.14788^*$ & $11.41718^*$ & $48.05146$ & $41.27006$ \\ \hline
$7$ & $1.130503$ & $10.34709^*$ & $49.88452$ & $25.84385^*$ & $42.7708$ & $36.8145$ \\ \hline
$8$ & $1.89275$ & $\textbf{5.027514}^*$ & $18.9799^*$ & $19.09505^*$ & $29.29209$ & $42.07779$ \\ \hline
$9$ & $0.490227$ & $\textbf{0.780852}^*$ & $6.021915^*$ & $27.74487$ & $39.82303$ & $46.06432$ \\ \hline
$10$ & $1.574057$ & $\textbf{2.557472}^*$ & $28.83701$ & $23.89729^*$ & $39.1091$ & $24.26693^*$ \\ \hline
$11$ & $3.910336$ & $11.32792^*$ & $15.90477^*$ & $7.477256^*$ & $40.16677$ & $27.36976$ \\ \hline
$12$ & $0.790672$ & $38.18567$ & $32.92128$ & $45.48335$ & $59.87374$ & $36.48986$ \\ \hline
\end{tabular}
\end{table} \\
Shah et al. $(1998)$ declared that all batches have unwanted changes in the amount of dissolution, and the dissolved values of batch $4$ in $30$ minutes are only dissimilar to the reference data. The maximum entropy reflects this reality via different $T^2$-Hotelling values in Table \ref{T2M}. Without any ambiguity, it is clear that all batches except batch $4$ are too different from the reference assortment. Although all tablets of batch $4$ are out of control based on $UCL_{ME}$, they are not as large as other batch values. So, the maximum entropy method detects undesirable changes concerning the reference group as well. Nine tablets in batch $4$ are in-control according to $UCL_F$, which means that the traditional way based on Fisher distribution is unable to realize the available changes in the data. The result of the linear regression in Table \ref{T2L} is quite different. This method does not completely detect changes in batch $1$, and some in-control tablets based on $UCL_{LR}$ can be seen in bold there. Also, $29$ in-control samples exist related to $UCL_F$ pointed by stars. Moreover, most of the tablets in batch $4$ are out of control that is far from the fact. So, decision correspondence to the linear regression method has some shortcomings. Thus, the power of maximum entropy is obvious and clear in this difficult situation, and other methods such as linear regression and Fisher distribution are unable to show the post-change of the data set in batches. \\
There is some other information from Shah et al. $(1998)$. The dissolved differences of batch $2$ are $15\%$ more than the reference group at $30$ minutes, but the differences reduced to less than $8\%$ at $60$, $90$, and $180$ minutes. In Tables \ref{WithoutM} and \ref{WithoutL}, they are compared in the absence of the column related to minute $30$. Batch $4$ at $30$ minutes is different from the reference group more than $12\%$, and the differences between this batch and the reference group ignore in the absence of $30$ minutes information and become similar. The reference $T^2$-Hotelling is provided in the tables to make a reliable comparison in the absence of the $30$ minutes column. The upper control limits at the confidence level $0.9973$ for this case are:
\begin{displaymath}
\left\{\begin{array}{ll}
UCL_{F}=26.97728, \\
UCL_{ME}=7.894543, \\
UCL_{LR}=7.232845. \\
\end{array}\right.
\end{displaymath}
\begin{table}[!h]
\centering
\caption{$T^2_{ME}$-Hotelling of reference and some batches are presented in the absence of some information which are mentions on the top of columns.}\label{WithoutM}
\begin{tabular}{|p{0.6in}||p{0.7in}|p{0.55in}|p{0.6in}||p{0.6in}|p{0.55in}||p{0.6in}|p{0.55in}|} \hline
Tablet numbers & Reference except $30$ min & Batch $2$ except $30$ min & Batch $4$ except $30$ min & Reference except $90$ min & Batch $3$ except $90$ min & Reference except $60$ min & Batch $5$ except $60$ min \\ \hline
$1$ & $0.85972098$ & $32054088$ & $0.859721$ & $0.6462575$ & $11342273$ & $1.0305838$ & $7100891$ \\ \hline
$2$ & $0.09665851$ & $31966736$ & $0.096659$ & $0.6621264$ & $11326617$ & $0.3331313$ & $7108760$ \\ \hline
$3$ & $1.07377578$ & $32042515$ & $1.073776$ & $0.7646651$ & $11329719$ & $0.8196994$ & $7094211$ \\ \hline
$4$ & $7.93191591$ & $32003090$ & $7.931916$ & $0.9894897$ & $11320807$ & $0.7515795$ & $7094148$ \\ \hline
$5$ & $6.67355851$ & $32066905$ & $6.673559$ & $6.4828865$ & $11335365$ & $6.5217488$ & $7095663$ \\ \hline
$6$ & $1.50996478$ & $32024502$ & $1.509965$ & $0.6737853$ & $11318497$ & $0.6987794$ & $7099653$ \\ \hline
$7$ & $0.4792845$ & $31984663$ & $0.479285$ & $0.9743507$ & $11343690$ & $1.5170518$ & $7102175$ \\ \hline
$8$ & $2.32876957$ & $32042369$ & $2.32877$ & $3.0835605$ & $11334122$ & $2.8849953$ & $7102262$ \\ \hline
$9$ & $0.36990082$ & $32054088$ & $0.369901$ & $0.3693712$ & $11341106$ & $0.4142522$ & $7097997$ \\ \hline
$10$ & $0.24985908$ & $32041637$ & $0.249859$ & $1.3754771$ & $11338460$ & $1.6259235$ & $7104667$ \\ \hline
$11$ & $5.78426564$ & $32055156$ & $5.784266$ & $3.9374103$ & $11330098$ & $2.4341753$ & $7100702$ \\ \hline
$12$ & $4.92606407$ & $32031147$ & $4.926064$ & $1.024471$ & $11331677$ & $0.8916272$ & $7102083$ \\ \hline
\end{tabular}
\end{table}
\begin{table}[!h]
\centering
\caption{$T^2_{LR}$-Hotelling of reference and some batches are presented in the absence of some information which are mentions on the top of columns.}\label{WithoutL}
\begin{tabular}{|p{0.6in}||p{0.6in}|p{0.6in}|p{0.55in}||p{0.6in}|p{0.6in}||p{0.6in}|p{0.6in}|} \hline
Tablet numbers & Reference except $30$ min & Batch $2$ except $30$ min & Batch $4$ except $30$ min & Reference except $90$ min & Batch $3$ except $90$ min & Reference except $60$ min & Batch $5$ except $60$ min \\ \hline
$1$ & $0.781326$ & $19.479468$ & $0.781326$ & $0.745188$ & $18.700255$ & $0.810631$ & $9.814556$ \\ \hline
$2$ & $0.104088$ & $31.607281$ & $0.104088$ & $0.617784$ & $12.67966$ & $0.414235$ & $14.40213$ \\ \hline
$3$ & $1.064175$ & $10.548568$ & $1.064175$ & $0.814136$ & $15.635688$ & $1.0593$ & $23.689262$ \\ \hline
$4$ & $7.2799$ & $14.664441$ & $7.2799$ & $0.340515$ & $29.588615$ & $0.238747$ & $19.500985$ \\ \hline
$5$ & $	5.695556$ & $31.995564$ & $5.695556$ & $5.938307$ & $19.407948$ & $5.934547$ & $8.270712$ \\ \hline
$6$ & $1.456388$ & $10.164522$ & $1.456388$ & $0.946578$ & $8.488775$ & $1.051532$ & $15.83993$ \\ \hline
$7$ & $0.34863$ & $17.763766$ & $0.34863$ & $1.122464$ & $25.023183$ & $1.404402$ & $14.090554$ \\ \hline
$8$ & $2.090627$ & $16.018051$ & $2.090627$ & $2.987033$ & $17.192682$ & $2.456116$ & $17.011613$ \\ \hline
$9$ & $0.504812$ & $19.479468$ & $0.504812$ & $0.435368$ & $23.507351$ & $0.547548$ & $18.129639$ \\ \hline
$10$ & $0.335571$ & $9.787093$ & $0.335571$ & $1.513262$ & $22.448421$ & $1.631706$ & $8.316706$ \\ \hline
$11$ & $5.169304$ & $21.786655$ & $5.169304$ & $3.465189$ & $6.065381$ & $2.042524$ & $6.870972$ \\ \hline
$12$ & $3.794569$ & $10.815753$ & $3.794569$ & $0.593449$ & $38.089939$ & $0.337266$ & $18.874469$ \\ \hline
\end{tabular}
\end{table} \\ \\
According to Table \ref{WithoutM}, the $T^2$-Hotelling statistics values are reduced related to batch $2$ and $4$, which means their distances are diminished from the reference group. Batch $2$ is still out of control, and the maximum entropy reflects it obviously, but batch $4$ becomes similar to the reference. The conclusion of the linear regression method shown in Table \ref{WithoutL} is quite the same. All tablets are out of control for batch $2$, and almost all of them for batch $4$ are in-control. Shah et al. $(1998)$ expressed about batch $3$ that the differences are more than $12\%$ at $90$ minutes between this batch and corresponding reference group information, and the differences become less than $10\%$ in the absence of data for $90$ minutes. The related $T^2$-Hotelling is in Tables \ref{WithoutM} and \ref{WithoutL} for when the data of $90$ minutes are omitted from the data set. Also, the related amounts of reference group are available the tables, and the corresponding upper control limits at the confidence level $0.9973$ are:
\begin{displaymath}
\left\{\begin{array}{ll}
UCL_{F}=26.97728, \\
UCL_{ME}=6.407286, \\
UCL_{LR}=5.864856. \\
\end{array}\right.
\end{displaymath}
The similarity of reference and batch $3$ in the absence of $90$ minutes is clearly rejected in the maximum entropy method. Although the $T^2$-Hotelling values are too high in this situation, they are less than the values calculated in Table \ref{T2M}. The result for the regression method is the same, and all dissolutions are out of control that declares that batch $3$ is not equivalent to the reference. The inference can be made for the last batch. Shah et al. $(1998)$ gave some further information about this batch as well. The difference between the batch and reference is more than $17\%$ for dissolved amounts of tablets at $60$ minutes. However, it is less than $10\%$ when the $60$ minutes of data are absent. In this case, the $T^2$-Hotelling statistic values are additionally calculated. The corresponding upper control limits at the significant level $0.0.0027$ are as following:
\begin{displaymath}
\left\{\begin{array}{ll}
UCL_{F}=26.97728, \\
UCL_{ME}=6.413737, \\
UCL_{LR}=5.831237. \\
\end{array}\right.
\end{displaymath}
The related information is in Tables \ref{WithoutM} and \ref{WithoutL}. In both methods, the similarity between batch $5$, and the reference is denied in the absence of $60$ minutes of data. Although all $T^2$-Hotelling amounts of dissolutions are less than the amounts in Tables \ref{T2M} and \ref{T2L}, and they show the existence differences, the maximum entropy decisively showed dissimilarity. The superiority of the maximum entropy principle was shown in this real situation of the profile dataset from the pharmaceutical production process.

\section{\bf Conclusion}
Usually, in statistical quality control, the purpose is to find suitable statistical limits to check processes during the time to keep them in control, or to detect their shifts as soon as possible in order not to waste money and time. So many methods are used for this aim. Most of them are based on process means, and the control charts monitor them. Up to our knowledge, they are not sufficient enough to detect different kinds of shifts. Therefore, in this paper, we would like to present a profile investigation of change points according to its coefficients. Our focus was on the simple linear profile, which looks like simple linear regression. Two methods of estimating are presented here. The first one is based on the maximum entropy principle compared with the second method, which is according to the linear regression. We use a $ T^2 $-Hotelling statistic to reduce the profile coefficient vector. \\
Generally, we would like to see shifts of processes by monitoring coefficients instead of means. To make a good comparison between their performance, we define three shifted models and calculate second errors $ \beta $, $ ARL_0 $, and $ ARL_1 $. plots based on second errors are drawn. We can conclude from simulation results that both of manners can detect changes in small shifts. But the method of maximum entropy has more performance. In the end, we provide a real data study to see the results in the methods. There was a change in the testing sample, and our purpose was to detect it. This kind of shift can not be detected based on mean control limits, but in the maximum entropy and the regression method based on profiles, it was detected easily. Thus, our methods were successful in checking the manufacturing process. In the last section, a pharmaceutical industry example was presented, which show the betterness of the maximum entropy principle as well. In this example, the regression procedure acts weak. Thus, we show their performance methods in three situations, simulated example,  semiconductor production, and pharmacy profiles. The ability of these two methods in the first two examples was somewhat the same, although the maximum entropy method noticed differences earlier. In the last example, the linear regression method did not correctly identify the dissimilarity, while the maximum entropy-based method did correctly identify the differences and similarities.\\

\section*{Data availability statement}
The data that support the findings of sections \ref{6} and \ref{7} are openly available in Zou et al. $(2007)$ at https://doi.org/10.1198/004017007000000164, and Shah et al. $(1998)$ at https://doi.org/10.1023/A:1011976615750, respectively. \\



\begin{thebibliography}{APA}

\bibitem[\protect\astroncite{Alfaro and Ortega}{2008}]{alfaro2008robust}
Alfaro, J.~L. and Ortega, J.~F. (2008).
\newblock A robust alternative to hotelling's t2 control chart using trimmed
  estimators.
\newblock {\em Quality and Reliability Engineering International},
  24(5):601--611.

\bibitem[\protect\astroncite{Aparisi}{1996}]{aparisi1996hotelling}
Aparisi, F. (1996).
\newblock Hotelling's t2 control chart with adaptive sample sizes.
\newblock {\em International Journal of Production Research},
  34(10):2853--2862.

\bibitem[\protect\astroncite{Aparisi and Haro}{2001}]{aparisi2001hotelling}
Aparisi, F. and Haro, C.~L. (2001).
\newblock Hotelling's t2 control chart with variable sampling intervals.
\newblock {\em International Journal of Production Research},
  39(14):3127--3140.

\bibitem[\protect\astroncite{Bhattacharya}{2006}]{bhattacharya2006maximum}
Bhattacharya, B. (2006).
\newblock Maximum entropy characterizations of the multivariate liouville
  distributions.
\newblock {\em Journal of Multivariate Analysis}, 97(6):1272--1283.

\bibitem[\protect\astroncite{Chakraborti et~al.}{2008}]{chakraborti2008phase}
Chakraborti, S., Human, S., and Graham, M. (2008).
\newblock Phase i statistical process control charts: an overview and some
  results.
\newblock {\em Quality Engineering}, 21(1):52--62.

\bibitem[\protect\astroncite{Champ and Woodall}{1987}]{champ1987exact}
Champ, C.~W. and Woodall, W.~H. (1987).
\newblock Exact results for shewhart control charts with supplementary runs
  rules.
\newblock {\em Technometrics}, 29(4):393--399.

\bibitem[\protect\astroncite{Chen}{1998}]{chen1998testing}
Chen, J. (1998).
\newblock Testing for a change point in linear regression models.
\newblock {\em Communications in Statistics-Theory and Methods},
  27(10):2481--2493.

\bibitem[\protect\astroncite{Chu and Satchell}{2009}]{chu2009most}
Chu, B. and Satchell, S. (2009).
\newblock The Most Entropic Canonical Copula With An Application to
  ‘style’investment.

\bibitem[\protect\astroncite{Costa et~al.}{2003}]{costa2003solutions}
Costa, J., Hero, A., and Vignat, C. (2003).
\newblock On solutions to multivariate maximum $\alpha$-entropy problems.
\newblock In {\em International Workshop on Energy Minimization Methods in
  Computer Vision and Pattern Recognition}, pages 211--226. Springer.

\bibitem[\protect\astroncite{Ebrahimi et~al.}{2008}]{ebrahimi2008multivariate}
Ebrahimi, N., Soofi, E.~S., Soyer, R., et~al. (2008).
\newblock Multivariate maximum entropy identification, transformation, and
  dependence.
\newblock {\em Journal of Multivariate Analysis}, 99(6):1217--1231.

\bibitem[\protect\astroncite{Fallah et~al.}{2019}]{fallah2019entropic}
Fallah, A., Mohtashami~Borzadaran, G.~R., and Sadeghpour~Gildeh, B. (2018).
\newblock An entropic structure in capability indices.
\newblock {\em Communications in Statistics-Theory and Methods},
        47(23), 5911-5921.

\bibitem[\protect\astroncite{Gupta et~al.}{2006}]{gupta2006performance}
Gupta, S., Montgomery, D., and Woodall, W. (2006).
\newblock Performance evaluation of two methods for online monitoring of linear
  calibration profiles.
\newblock {\em International Journal Of Production Research},
  44(10):1927--1942.

\bibitem[\protect\astroncite{Hawkins}{1989}]{hawkins1989ui}
Hawkins, D. (1989).
\newblock A ui approach to retrospective testing for shifting parameters in a
  linear model.
\newblock {\em Communications in Statistics-Theory and Methods},
  18(8):3117--3134.

\bibitem[\protect\astroncite{Hawkins}{1991}]{hawkins1991multivariate}
Hawkins, D.~M. (1991).
\newblock Multivariate quality control based on regression-adiusted variables.
\newblock {\em Technometrics}, 33(1):61--75.

\bibitem[\protect\astroncite{Hawkins}{1993}]{hawkins1993regression}
Hawkins, D.~M. (1993).
\newblock Regression adjustment for variables in multivariate quality control.
\newblock {\em Journal of Quality Technology}, 25(3):170--182.

\bibitem[\protect\astroncite{Haworth}{1996}]{haworth1996regression}
Haworth, D.~A. (1996).
\newblock Regression control charts to manage software maintenance.
\newblock {\em Journal of Software Maintenance: Research and Practice},
  8(1):35--48.

\bibitem[\protect\astroncite{Hillier}{1969}]{hillier1969x}
Hillier, F.~S. (1969).
\newblock X-and r-chart control limits based on a small number of subgroups.
\newblock {\em Journal of Quality Technology}, 1(1):17--26.

\bibitem[\protect\astroncite{Holbert}{1982}]{holbert1982bayesian}
Holbert, D. (1982).
\newblock A bayesian analysis of a switching linear model.
\newblock {\em Journal of Econometrics}, 19(1):77--87.

\bibitem[\protect\astroncite{Jaynes}{1957}]{jaynes1957information}
Jaynes, E.~T. (1957).
\newblock Information theory and statistical mechanics.
\newblock {\em Physical Review}, 106(4):620.

\bibitem[\protect\astroncite{Jones}{1976}]{jones1976multivariate}
Jones, R.~H. (1976).
\newblock Multivariate maximum entropy spectral analysis.
\newblock In {\em Appl. Time Ser. Anal. Symp.}

\bibitem[\protect\astroncite{Kagan et~al.}{1973}]{kagan1973extension}
Kagan, A., Linnik, Y.~V., and Rao, C. (1973).
\newblock Extension of darmois-skitcvic theorem to functions of random
  variables satisfying an addition theorem.
\newblock {\em Communications in Statistics-Theory and Methods}, 1(5):471--474.

\bibitem[\protect\astroncite{Kang and Albin}{2000}]{kang2000line}
Kang, L. and Albin, S.~L. (2000).
\newblock On-line monitoring when the process yields a linear profile.
\newblock {\em Journal of quality Technology}, 32(4):418--426.

\bibitem[\protect\astroncite{Kim}{1994}]{kim1994tests}
Kim, H.-J. (1994).
\newblock Tests for a change-point in linear regression.
\newblock {\em Lecture Notes-Monograph Series}, pages 170--176.

\bibitem[\protect\astroncite{Kim and Siegmund}{1989}]{kim1989likelihood}
Kim, H.-J. and Siegmund, D. (1989).
\newblock The likelihood ratio test for a change-point in simple linear
  regression.
\newblock {\em Biometrika}, 76(3):409--423.

\bibitem[\protect\astroncite{King}{1954}]{king1954probability}
King, E.~P. (1954).
\newblock Probability limits for the average chart when process standards are
  unspecified.
\newblock {\em Industrial Quality Control}, 10(6):62--64.

\bibitem[\protect\astroncite{Kopnov and Kanajev}{1994}]{kopnov1994optimal}
Kopnov, V. and Kanajev, E. (1994).
\newblock Optimal control limit for degradation process of a unit modelled as a
  markov chain.
\newblock {\em Reliability Engineering \& System Safety}, 43(1):29--35.

\bibitem[\protect\astroncite{Kouskoulas
  et~al.}{2004}]{kouskoulas2004computationally}
Kouskoulas, Y., Pierce, L.~E., and Ulaby, F.~T. (2004).
\newblock A computationally efficient multivariate maximum-entropy density
  estimation (mede) technique.
\newblock {\em IEEE Transactions On Geoscience And Remote Sensing},
  42(2):457--468.

\bibitem[\protect\astroncite{Krogh and Mitchison}{1995}]{krogh1995maximum}
Krogh, A. and Mitchison, G.~J. (1995).
\newblock Maximum entropy weighting of aligned sequences of proteins or dna.
\newblock In {\em ISMB}, volume~3, pages 215--221.

\bibitem[\protect\astroncite{Lawless et~al.}{1999}]{lawless1999analysis}
Lawless, J., Mackay, R., and Robinson, J. (1999).
\newblock Analysis of variation transmission in manufacturing processes—part i.
\newblock {\em Journal of Quality Technology}, 31(2):131--142.

\bibitem[\protect\astroncite{Ma et~al.}{2000}]{ma2000assessment}
Ma, M.-C., Wang, B.~B., Liu, J.-P., and Tsong, Y. (2000).
\newblock Assessment of similarity between dissolution profiles.
\newblock {\em Journal of Biopharmaceutical Statistics}, 10(2):229--249.

\bibitem[\protect\astroncite{Mahmoud et~al.}{2007}]{mahmoud2007change}
Mahmoud, M.~A., Parker, P.~A., Woodall, W.~H., and Hawkins, D.~M. (2007).
\newblock A change point method for linear profile data.
\newblock {\em Quality and Reliability Engineering International},
  23(2):247--268.

\bibitem[\protect\astroncite{Mahmoud and Woodall}{2004}]{mahmoud2004phase}
Mahmoud, M.~A. and Woodall, W.~H. (2004).
\newblock Phase i analysis of linear profiles with calibration applications.
\newblock {\em Technometrics}, 46(4):380--391.

\bibitem[\protect\astroncite{Mandel}{1969}]{mandel1969regression}
Mandel, B. (1969).
\newblock The regression control chart.
\newblock {\em Journal of Quality Technology}, 1(1):1--9.

\bibitem[\protect\astroncite{May et~al.}{1991}]{may1991statistical}
May, G.~S., Huang, J., and Spanos, C.~J. (1991).
\newblock Statistical experimental design in plasma etch modeling.
\newblock {\em IEEE Transactions on Semiconductor Manufacturing}, 4(2):83--98.

\bibitem[\protect\astroncite{Montgomery}{2007}]{montgomery2007introduction}
Montgomery, D.~C. (2007).
\newblock {\em Introduction to Statistical Quality Control}.
\newblock John Wiley \& Sons.

\bibitem[\protect\astroncite{Mortezanejad et~al.}{2019}]{mortezanejad2019joint}
Mortezanejad, S. A.~F., Borzadaran, G.~M., and Sadeghpour~Gildeh, B. (2019).
\newblock Joint dependence distribution of data set using optimizing tsallis
  copula entropy.
\newblock {\em Physica A: Statistical Mechanics and its Applications},
  533,121897.

\bibitem[\protect\astroncite{Neubauer}{1997}]{neubauer1997ewma}
Neubauer, A.~S. (1997).
\newblock The ewma control chart: properties and comparison with other
  quality-control procedures by computer simulation.
\newblock {\em Clinical Chemistry}, 43(4),594--601.

\bibitem[\protect\astroncite{Pougaza and
  Mohammad-Djafari}{2011}]{pougaza2011maximum}
Pougaza, D.-B. and Mohammad-Djafari, A. (2011).
\newblock Maximum entropies copulas.
\newblock In {\em AIP Conference Proceedings}, volume 1305, pages 329--336.
  AIP.

\bibitem[\protect\astroncite{Pougaza and
  Mohammad-Djafari}{2012}]{pougaza2012new}
Pougaza, D.-B. and Mohammad-Djafari, A. (2012).
\newblock New copulas obtained by maximizing tsallis or r{\'e}nyi entropies.
\newblock In {\em AIP Conference Proceedings 31st}, volume 1443, pages
  238--249. AIP.

\bibitem[\protect\astroncite{Prabhu and Runger}{1997}]{prabhu1997designing}
Prabhu, S.~S. and Runger, G.~C. (1997).
\newblock Designing a multivariate ewma control chart.
\newblock {\em Journal of Quality Technology}, 29(1):8--15.

\bibitem[\protect\astroncite{Quandt}{1958}]{quandt1958estimation}
Quandt, R.~E. (1958).
\newblock The estimation of the parameters of a linear regression system
  obeying two separate regimes.
\newblock {\em Journal Of The American Statistical Association},
  53(284):873--880.

\bibitem[\protect\astroncite{Quercia et~al.}{2012}]{quercia2012tracking}
Quercia, D., Ellis, J., Capra, L., and Crowcroft, J. (2012).
\newblock Tracking gross community happiness from tweets.
\newblock In {\em Proceedings Of The ACM 2012 Conference On Computer Supported
  Cooperative Work}, pages 965--968. ACM.

\bibitem[\protect\astroncite{Shah et~al.}{1998}]{shah1998vitro}
Shah, V.~P., Tsong, Y., Sathe, P., and Liu, J.-P. (1998).
\newblock In vitro dissolution profile comparison?statistics and analysis of
  the similarity factor, f2.
\newblock {\em Pharmaceutical Research}, 15(6):889--896.

\bibitem[\protect\astroncite{Shannon}{1948}]{shannon1948mathematical}
Shannon, C.~E. (1948).
\newblock A mathematical theory of communication.
\newblock {\em Bell System Technical Journal}, 27(3):379--423.

\bibitem[\protect\astroncite{Shore and Johnson}{1980}]{shore1980axiomatic}
Shore, J. and Johnson, R. (1980).
\newblock Axiomatic derivation of the principle of maximum entropy and the
  principle of minimum cross-entropy.
\newblock {\em IEEE Transactions On Information Theory}, 26(1):26--37.

\bibitem[\protect\astroncite{Smith and Livesey}{1992}]{smith1992maximum}
Smith, G. and Livesey, A. (1992).
\newblock Maximum entropy: A new approach to non-destructive deconvolution of
  depth profiles from angle-dependent xps.
\newblock {\em Surface And Interface Analysis}, 19(1-12):175--180.

\bibitem[\protect\astroncite{Sparks}{2000}]{sparks2000cusum}
Sparks, R.~S. (2000).
\newblock Cusum charts for signalling varying location shifts.
\newblock {\em Journal of Quality Technology}, 32(2):157--171.

\bibitem[\protect\astroncite{Urz{\'u}a}{1988}]{urzua1988class}
Urz{\'u}a, C. (1988).
\newblock A class of maximum-entropy multivariate distributions.
\newblock {\em Communications in Statistics-Theory and Methods},
  17(12):4039--4057.

\bibitem[\protect\astroncite{Wade and Woodall}{1993}]{wade1993review}
Wade, M.~R. and Woodall, W.~H. (1993).
\newblock A review and analysis of cause-selecting control charts.
\newblock {\em Journal of Quality Technology}, 25(3):161--169.

\bibitem[\protect\astroncite{Westgard et~al.}{1977}]{westgard1977combined}
Westgard, J., Groth, T., Aronsson, T., and De~Verdier, C. (1977).
\newblock Combined shewhart-cusum control chart for improved quality control in
  clinical chemistry.
\newblock {\em Clinical Chemistry}, 23(10):1881--1887.

\bibitem[\protect\astroncite{Woodall}{2007}]{woodall2007current}
Woodall, W.~H. (2007).
\newblock Current research on profile monitoring.
\newblock {\em Production}, 17(3):420--425.

\bibitem[\protect\astroncite{Woodall}{2017}]{woodall2017bridging}
Woodall, W.~H. (2017).
\newblock Bridging the gap between theory and practice in basic statistical
  process monitoring.
\newblock {\em Quality Engineering}, 29(1):2--15.

\bibitem[\protect\astroncite{Woodall et~al.}{2004}]{woodall2004using}
Woodall, W.~H., Spitzner, D.~J., Montgomery, D.~C., and Gupta, S. (2004).
\newblock Using control charts to monitor process and product quality profiles.
\newblock {\em Journal of Quality Technology}, 36(3):309--320.

\bibitem[\protect\astroncite{Wu}{2003}]{wu2003calculation}
Wu, X. (2003).
\newblock Calculation of maximum entropy densities with application to income
  distribution.
\newblock {\em Journal of Econometrics}, 115(2):347--354.

\bibitem[\protect\astroncite{Wu and Perloff}{2007}]{wu2007gmm}
Wu, X. and Perloff, J.~M. (2007).
\newblock Gmm estimation of a maximum entropy distribution with interval data.
\newblock {\em Journal of Econometrics}, 138(2):532--546.

\bibitem[\protect\astroncite{Wu et~al.}{2009}]{wu2009enhanced}
Wu, Z., Jiao, J., Yang, M., Liu, Y., and Wang, Z. (2009).
\newblock An enhanced adaptive cusum control chart.
\newblock {\em IIE Transactions}, 41(7):642--653.

\bibitem[\protect\astroncite{Xie et~al.}{2000}]{xie2000exponential}
Xie, M., Kong, H., and Goh, T. (2000).
\newblock Exponential approximation for maintained weibull distributed
  component.
\newblock {\em Journal of Quality in Maintenance Engineering}, 6(4):260--269.

\bibitem[\protect\astroncite{Yang and Hillier}{1970}]{yang1970mean}
Yang, C.-H. and Hillier, F.~S. (1970).
\newblock Mean and variance control chart limits based on a small number of
  subgroups.
\newblock {\em Journal of Quality Technology}, 2(1):9--16.

\bibitem[\protect\astroncite{Zellner and
  Highfield}{1988}]{zellner1988calculation}
Zellner, A. and Highfield, R.~A. (1988).
\newblock Calculation of maximum entropy distributions and approximation of
  marginalposterior distributions.
\newblock {\em Journal of Econometrics}, 37(2):195--209.

\bibitem[\protect\astroncite{Zhang}{1992}]{zhang1992cause}
Zhang, G. (1992).
\newblock {\em Cause-selecting control chart and diagnosis: Theory and
  practise}.
\newblock {\em Aarhus School of Business.}

\bibitem[\protect\astroncite{Zou and Tsung}{2011}]{zou2011multivariate}
Zou, C. and Tsung, F. (2011).
\newblock A multivariate sign ewma control chart.
\newblock {\em Technometrics}, 53(1):84--97.

\bibitem[\protect\astroncite{Zou et~al.}{2007}]{zou2007monitoring}
Zou, C., Tsung, F., and Wang, Z. (2007).
\newblock Monitoring general linear profiles using multivariate exponentially
  weighted moving average schemes.
\newblock {\em Technometrics}, 49(4):395--408.

\end{thebibliography}

\end{document}